\documentclass[aps,prb,floatfix,amsmath,amssymb,preprint,eqsecnum,nofootinbib]{revtex4-1}
\usepackage{graphicx}
\usepackage{dcolumn}
\usepackage{bm}
\usepackage{setspace}   
\usepackage{color}

\usepackage{setspace} 
\newcommand{\beq}{\begin{equation}}
\newcommand{\eeq}{\end{equation}}
\newcommand{\ba}{\begin{array}{ccc}}
\newcommand{\ea}{\end{array}}
\newcommand{\nn}{\nonumber}
 \renewcommand{\d}{\partial}
\def\bea{\begin{eqnarray}}
\def\eea{\end{eqnarray}}

\def\<{\langle}
\def\>{\rangle}
\usepackage{amsmath}
\usepackage{amssymb}
\begin{document}
\preprint{arXiv:1001.1153}
\title{Quantum phase transitions of metals in two spatial dimensions:\\ I. Ising-nematic order}

\author{Max A. Metlitski}
\affiliation{Department of Physics, Harvard University, Cambridge MA
02138}

\author{Subir Sachdev}
\affiliation{Department of Physics, Harvard University, Cambridge MA
02138}

\date{January 6, 2010\\
\vspace{1.6in}}
\begin{abstract}
We present a renormalization group theory for the onset of Ising-nematic order
in a Fermi liquid in two spatial dimensions. This is a quantum phase transition, driven by electron 
interactions, which
spontaneously reduces the point-group symmetry from square to rectangular.
The critical point is described by an infinite set of 2+1 dimensional local field theories, labeled by points on the Fermi surface.
Each field theory 
contains a real scalar field representing the Ising order parameter, and fermionic fields
representing a time-reversed pair of patches on the Fermi surface. 
We demonstrate that the field theories obey compatibility constraints required by our redundant representation
of the underlying degrees of freedom.
Scaling forms for the response
functions are proposed, and supported by computations up to three loops. Extensions of our results
to other transitions of two-dimensional Fermi liquids 
with broken point-group and/or time-reversal symmetry are noted. Our results extend also to the problem
of a Fermi surface coupled to a U(1) gauge field.
\end{abstract}

\maketitle

\section{Introduction}
\label{sec:intro}

A number of recent experiments \cite{ando02,hinkov08a,kohsaka07,taill10b} 
have noted the presence of Ising-nematic order
in the enigmatic normal state of the cuprate superconductors. This order is associated with electronic
correlations which spontaneously break the square lattice symmetry to that of a rectangular
lattice: {\em i.e.\/} the symmetry of 90$^\circ$ rotations is lost, and the $x$ and $y$ directions become
inequivalent. This broken symmetry is associated with an Ising order parameter, which we will
represent below by a real scalar field $\phi$.

Of particular interest are recent experiments on the anisotropy of the Nernst signal \cite{taill10b}
in YBa$_2$Cu$_3$O$_y$,
which indicate that the Ising-nematic order has its onset at the temperature $T=T^\ast$,
which also marks the boundary between the `pseudogap' region and the `strange metal'.
These results call for the theory of the quantum phase transition  involving Ising-nematic
ordering in a Fermi liquid metal. Such a quantum critical point would play an important role
in the theory of the strange metal. The metallic Ising-nematic critical point
is also of importance in experiments \cite{borzi07} on Sr$_3$Ru$_2$O$_7$, where the observations of resistance
anisotropies have demonstrated spontaneous Ising-nematic ordering. Finally, there are clear
indications of Ising-nematic order driven by electron correlations in the pnictides. \cite{pnic1,pnic2,pnic3,pnic4}

One approach to the Ising-nematic ordering is to take a liquid-crystalline perspective \cite{KFE98},
and view it among a class of phases with broken square lattice symmetry \cite{zaanen,kiv_rmp,ss_rmp,mv_ap}. 
Ising nematic phases are also a generic feature of frustrated and doped antiferromagnets, 
because the Ising-nematic order survives after antiferromagnetism (at wavevectors $\neq (\pi, \pi)$) has been disrupted by
thermal \cite{ccl,luca} or quantum \cite{rs2,rs2l} fluctuations.

A complementary
point of view \cite{YK00,HM00,OKF01,MRA03,KKC03,YOM05,DM06,RPC06,LFBFO06,LF07,JSKM08,ZWG09,MC09} is to start from the Fermi liquid with perfect square lattice symmetry and look for the Pomeranchuk
instability of Landau's Fermi liquid theory in the angular momentum $\ell = 2$ channel.
Almost all of these works rely on the perspective of Hertz \cite{vrmp}, in which the electrons are integrated out
to yield a Landau-damped effective action for the scalar order parameter $\phi$; the low energy particle-hole excitations
near the Fermi surface lead to long-range interactions in the action for $\phi$. However, this procedure of successive integration
of fermionic and then bosonic degrees of freedom is clearly dangerous. A systematic renormalization group analysis requires
that all excitations at a given energy scale be treated together. Consequently, a complete scaling analysis of the
Ising nematic critical point is lacking: such an analysis should be based on a local field theory, and provide a scheme for
computing the scaling dimensions of all perturbations of the critical point.

We can also consider the onset of Ising-nematic order in a superconductor, rather than in a Fermi liquid.
In a $s$-wave superconductor, the fermionic excitations are fully gapped, and so the theory for $\phi$ has no long-range
interactions: consequently the transition is in the universality class of the 2+1 dimensional pure Ising model.
A $d$-wave superconductor does have gapless fermionic excitations at special `nodal points' in the Brillouin
zone, and these nodal fermions do modify the universality of the transition away from pure Ising \cite{vojtaprl,vojtaijm}.
A fairly complete understanding of the Ising-nematic transition in $d$-wave superconductors has been reached in
recent work \cite{kim,huh} using a large-$N$ expansion, where $N$ is the number of fermion components.

This paper provides a scaling theory of the Ising-nematic quantum critical point in two-dimensional metals,
satisfying the requirements stated above. Our theory builds upon the work in the $d$-wave superconductor \cite{kim,huh},
and also on advances by Polchinski\cite{Polchinski}, Altshuler, Ioffe, and Millis\cite{ALTSHULER}, 
and Sung-Sik Lee\cite{sslee2,SSLee} 
on a closely-related problem: the dynamics of a Fermi
surface with the fermions coupled minimally to a U(1) gauge field. 

We focus on a pair of time-reversed patches
on the Fermi surface and describe their vicinity by a local 2+1 dimensional field theory. In principle, there are separate
critical theories for each pair of time-reversed points on the Fermi surface, as is also the case in the
Fermi surface `bosonization' methods.\cite{Luther,Marston,KwonMarston,haldane,NetoFradkin,LF07,LFBFO06} 
However, a key difference from the latter methods is that each Fermi surface
point is associated with a 2+1 dimensional theory, and not a 1+1 a dimensional theory. 
This means that there is a redundancy
in our description, and sowing the theories together is not trivial: we show in Section~\ref{sec:rot} how
this is done in a consistent manner.

Apart from their application to the Ising-nematic transition of interest,
simple extensions of our results apply also to the U(1) gauge field case, and 
to other symmetry breaking transitions in Fermi liquids involving order parameters
which carry momentum $\vec{Q} = 0$. We will describe these cases in Section~\ref{sec:model} below,
and briefly indicate the needed extensions in the body of the paper.

Transitions with order parameters which carry momentum $\vec{Q} \neq 0$ lead to 
different field theories, which will be described in a subsequent paper.\cite{part2}

After a discussion of the one loop results in Section~\ref{sec:oneloop}, we present our main scaling
analysis in Section~\ref{sec:sr}. This includes a discussion of Ward identities which strongly
constrain the structure of renormalization group flow. Finally, explicit three loop computations appear in Section~\ref{sec:aetl} and Appendix \ref{app:comp}.

\section{The model}\label{sec:model}

We consider quantum phase transitions in metals of electrons $c_\sigma$ ($\sigma = \uparrow, \downarrow$), involving
an onset of a real order parameter $\phi(x)$ at wave-vector $\vec{Q} = 0$. The order parameter
is taken to have the same transformation properties under lattice symmetries and time reversal as,
\beq O(\vec{x}) = \frac{1}{V} \sum_{\vec{q}} \sum_{\vec{k} \sigma} d_{\vec{k} \sigma} c^\dagger_{\vec{k}-\vec{q}/2, \sigma}  c_{\vec{k}+\vec{q}/2, \sigma} e^{i \vec{q} \cdot \vec{x}}
\label{defO}
\eeq
For definiteness, we consider a system on a square lattice. Then, $\phi$ can describe the following patterns of symmetry breaking:
\begin{enumerate}
\item Breaking of the point-group symmetry with $d_{\vec{k} \uparrow} = d_{\vec{k} \downarrow}$ and $d_{\vec{k} \sigma} = d_{- \vec{k} \sigma}$. In these cases
$d_{\vec{k}}$ has either $d_{x^2 - y^2}$, $d_{xy}$,
or $g$-wave symmetry. The Ising-nematic transition of most interest to us here corresponds to the $d_{x^2-y^2}$ or $d_{xy}$ cases.
These cases all belong to one-dimensional representations of the square lattice point group, and 
we will argue that these transitions are all in the same universality class. 
\item Breaking of time-reversal and point-group symmetry with $d_{\vec{k} \uparrow} = d_{\vec{k} \downarrow}$ 
and $d_{\vec{k} \sigma} = -d_{- \vec{k} \sigma}$. In this case $d_{\vec{k}}$ transforms under the two-dimensional $p$-wave
representation, and so requires a two component order parameter $\vec{\phi} =  (\phi_x , \phi_y)$. 
We will not consider the two-component case explicitly, but 
our results have an immediate generalization to this transition. This case corresponds to the ``circulating current'' order parameters proposed
by Simon and Varma \cite{varma}, as was argued in Refs.~\onlinecite{vojtaprl,berg}.
\item Breaking of spin-inversion symmetry with 
$d_{\vec{k} \uparrow} = - d_{\vec{k} \downarrow}$. In this case, $d_{\vec{k}}$ can have either $s$-wave symmetry (Ising ferromagnet), $d$-wave symmetry (Ising spin-nematic) or $g$-wave symmetry. Unlike transitions i) and ii), which respect the full $SU(2)$ spin rotation symmetry, in the present case we assume this symmetry is explicitly broken to a $U(1)$ ``easy axis" subgroup. 
\end{enumerate}
Notice that in all cases, there is a $Z_2$ symmetry (either $\pi/2$ rotation, reflection or time-reversal) under which $\phi \to -\phi$. 

Apart from the above symmetry breaking cases, we will also consider the problem of a Fermi surface
minimally coupled to a U(1) gauge field \cite{SSLee,sslee2,Polchinski,HALPERIN,KIM94,NAYAK,ALTSHULER,ybkim2,stern1,stern2,read,rshankar,hermele,senthil}.
This case is similar to case 2 above, as we describe below Eq.~(\ref{Ldim}). 
Such models arise in theories \cite{hermele,senthil} of certain U(1) spin liquid phases in which $c_\sigma$ describe the fermionic spinons.
We will therefore refer to this model as the ``spin-liquid'' case below. The same theory also
describes \cite{rkk2,su2,qi} ``algebraic charge liquids'' in which case the $c_\sigma$ are 
spinless, charge $-e$ fermions, and $\sigma$ represents the charge of the fermion under the emergent U(1) gauge field;
we will not refer to this case explicitly below.

Given the order parameter in Eq.~(\ref{defO}), 
we may write down an effective spacetime Lagrangian describing the interactions of the order parameter $\phi$ with the fermions as,
\beq L =  c^\dagger_{\sigma} \Bigl( \d_{\tau} + \epsilon(- i \nabla) \Bigr) c_{\sigma}  - O(x) \phi(x)
+ \frac{1}{2} (\nabla \phi)^2 + \frac{r_0}{2} \phi^2 \label{ssL} \eeq
Here, we have added by hand a gradient term and a mass for the bosonic mode $\phi$. Such terms will be generated automatically after integrating out the high-energy fermions. The absence of higher order terms in $\phi$ and gradients of $\phi$ will be justified below.

The Lagrangian $L$ in Eq.~(\ref{ssL}) is not yet in a form suitable for our analysis of quantum criticality. The main
point is that the fermion spectrum $\epsilon (\vec{k})$ has zeros along the entire Fermi surface of large momenta $\vec{k}$:
so, as is well known, we are not in a position to make a low momentum expansion needed for a field theory. One strategy is to use
the Hertz approach \cite{vrmp} of integrating out all the $c$ fermions to obtain a non-local effective action
for the order parameter $\phi$. The latter is singular only at small momenta $\vec{q}$ and $\omega$, and so it is then at least permissible to make a low momentum and frequency expansion. However, the terms in the effective for $\phi$ turn 
out to be highly singular as $\vec{q} \rightarrow 0$ (see Ref.~\onlinecite{RPC06} and Appendix \ref{app:decoupling}). Moreover, in $d=2$, the strength of the singularity increases with increasing
powers of $\phi$ in the effective action. 
The situation now seems hopeless, but progress becomes possible after a key observation:
the leading singularities in the $\phi$ effective action appear only when all the $\phi$ fields have their momenta nearly collinear
to each other, as is explained in Appendix~\ref{app:decoupling}, and as
will become clear from the structure of our analysis below (by nearly collinear we mean that the angle $\theta$ between the momenta
is of order $\theta \sim |\vec{q}|/k_F$). 
In other words, if we are interested only in leading critical behavior, $\phi$ fields with non-collinear
momenta effectively decouple from each other. The couplings between $\phi$ fields with non-collinear momenta are 
then irrelevant corrections to the critical theory. The argument supporting this statement is presented in Appendix \ref{app:decoupling}. More generally, consider an $n$-point function
\begin{displaymath}
\left \langle \phi ( \vec{q}_1 ) \phi (\vec{q}_2 ) \phi (\vec{q}_3 ) \ldots \phi (\vec{q}_n ) \right\rangle.
\end{displaymath}
In a Gaussian theory for $\phi$, which is the claim of Hertz \cite{vrmp}, such a correlator would decouple
into products over pairs of momenta which sum to zero. However, such a decoupling is too drastic: rather,
the decoupling is only over sets of momenta which are collinear with each other, so that the leading critical singularity
of the above correlator
takes the form
\begin{displaymath}
\prod_a 
\left \langle \phi ( \vec{q}_{a1} ) \phi (\vec{q}_{a2} ) \ldots \right\rangle.
\end{displaymath}
Here all the momenta $\vec{q}_{ai}$ in a group $Q_a$ are collinear
to each other, while being non-collinear to momenta in groups $Q_b$ with $b \neq a$.
We can therefore limit ourselves to $\phi$ fields with momenta along a fixed direction $\vec{q}$. 
We will now argue that for each such direction $\vec{q}$, there is a sensible and powerful continuum limit of Eq.~(\ref{ssL}).

\begin{figure}[ht]
\includegraphics*[width=3.5in]{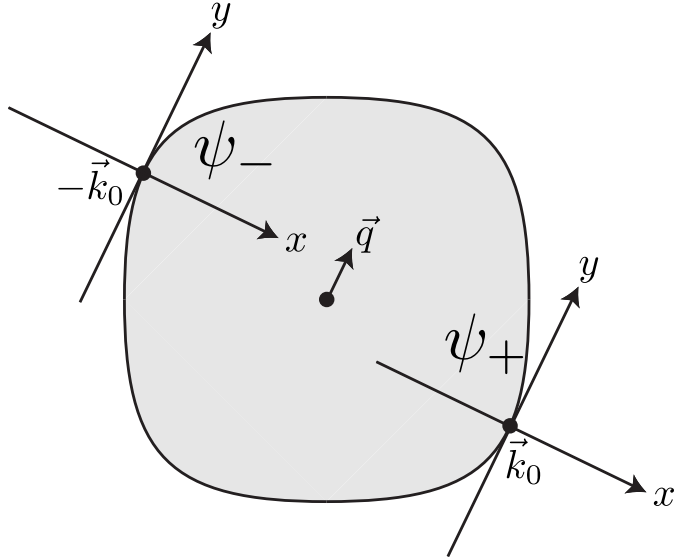}
\caption{The shaded region represents the occupied states inside a Fermi surface.
Fluctuations of the order parameter $\phi$ at wavevectors parallel to $\vec{q}$ couple most strongly to fermions near the Fermi 
surface points $\pm \vec{k}_0$. These fermions are denoted $\psi_\pm$.}
\label{fig:fs}
\end{figure}

It is now clear that we may restrict our search for a field theory to that describing the singularities in the $\phi$ correlations for a {\em single\/}
group of collinear momenta $Q_a$. So let us pick a direction $\vec{q}$ for $\phi$.
It is believed that a bosonic mode with momentum $\vec{q}$ interacts most strongly with the patches of the Fermi-surface to which it is tangent \cite{SSLee,sslee2,Polchinski,ALTSHULER}. Assuming that only a single Fermi surface is present, for each $\vec{q}$ there will be two such points with opposite Fermi-momenta $\vec{k}_0$ and $-\vec{k}_0$, see Fig.~\ref{fig:fs}.
We will denote fermions at these momenta as $\psi_+$ and $\psi_-$: 
\beq \psi_{+\sigma} (\vec{k} ) = c_{\vec{k}_0 + \vec{k}, \sigma}~~,~~\psi_{-\sigma} (\vec{k} ) = c_{-\vec{k}_0 + \vec{k}, \sigma}. \label{psic} \eeq
We choose coordinate vectors  $\hat{x}$ and $\hat{y}$ to be respectively perpendicular and parallel to $\vec{q}$. Then, expanding the fermion energy near $\vec{k}_0$ and $-\vec{k}_0$, the needed, low energy, continuum 
Lagrangian becomes
\bea L_{k_0} &=& \psi^\dagger_{+\sigma} \left( \d_{\tau} - i v_F \d_x - \frac{1}{2m} \d^2_y \right) \psi_{+ \sigma}  + \psi^\dagger_{-\sigma} \left(\d_{\tau} + i v_F \d_x - \frac{1}{2m} \d^2_y \right) \psi_{- \sigma} \nonumber \\  &-& d_{+\sigma} \, \phi \, \psi^\dagger_{+\sigma} \psi_{+\sigma} - d_{-\sigma} \, \phi \, \psi^\dagger_{-\sigma} \psi_{-\sigma}
+ \frac{1}{2} (\d_y \phi)^2 + \frac{r_0}{2} \phi^2 \label{Ldim}\eea
Here $v_F$ and $m$ are the Fermi velocity and the band mass at $k_0$, while $d_{\pm \sigma} = d_{\pm k_0 \sigma}$, and
we have added a subscript $k_0$ to $L$ emphasize that this is the Lagrangian for the patch near $\pm \vec{k}_0$.

We should emphasize here that all the fields in Eq.~(\ref{Ldim}) are 2+1 dimensional quantum fields, with full
dependence upon $x$, $y$, and $\tau$ {\em i.e.\/} the fields are $\phi(x, y, \tau)$ and $\psi_{\pm \sigma} (x, y, \tau)$.
In principle, we should also add a term $(\partial_x \phi)^2$ to Eq.~(\ref{Ldim}); however, we omit it at the outset
because it will later be seen to be irrelevant near criticality.
Further, because of this full dependence on $x$, and $y$, the fermion fields $\psi_{\pm \sigma}$ describe an
extended patch of the Fermi surface near the points $\pm \vec{k}_0$, and not just the two points $\pm \vec{k}_0$. 
We place some finite cutoff $\Lambda$ on the size of this patch, and will be interested in the scaling behavior at momenta
much smaller than this cutoff.

We now discuss the structure of the couplings $d_{\pm \sigma}$ in Eq.~(\ref{Ldim}).
For the transitions in $s$, $d$ and $g$ channels in case 1 above $d_{+\sigma} = d_{-\sigma}$ by inversion symmetry,
and $d_{\pm\sigma}$ is $\sigma$ independent. For case 2, we have $d_{+\sigma} = -d_{-\sigma}$ and also $\sigma$ independent,
although the fermions now couple to a projection of the two component order parameter $\vec{\phi} \cdot \vec{d}$, while the bosonic gradient term generally involves both components of the order parameter.  The spin liquid case also has
$d_{+\sigma} = -d_{-\sigma}$ and $\sigma$ independent, and
$\phi$ is associated with the transverse component of the spatial gauge field in the Coulomb gauge \cite{SSLee,sslee2,Polchinski,ALTSHULER}; moreover the spin-liquid has $r=0$ by gauge invariance. Finally, the Ising ferromagnet case 3 has $d_{+\sigma}=d_{-\sigma}$ and $d_{\pm \uparrow} = - d_{\pm \downarrow}$.

We note that for transitions in non-zero angular momentum channels, the coupling $d$ vanishes along certain axes in the Brillouin zone. The intersections of these axes with the Fermi surface are known as cold-spots, as the fermion coupling to the order parameter at these points involves additional derivatives and is much weaker. The scaling theory that follows only describes the Fermi surface away from cold spots.

It is convenient to rescale coordinates and fields in (\ref{Ldim}), $x = (2 m v_F)^{-1} \tilde{x}$, $\psi = v_F^{-1/2} \tilde{\psi}$, $\phi = \frac{1}{2 m |d|} \tilde{\phi}$. We drop the tildes in what follows. Then,
\bea L &=& \psi^\dagger_{+ \sigma} \Bigl( \eta \d_{\tau} - i  \d_x -  \d^2_y \Bigr) \psi_{+ \sigma}  
+ \psi^\dagger_{-\sigma} \Bigl( \eta \d_{\tau} + i  \d_x -  \d^2_y \Bigr) \psi_{- \sigma} \nonumber \\
&~&~~~ - \lambda_{+\sigma} \, \phi \, \psi^\dagger_{+ \sigma} \psi_{+ \sigma}- \lambda_{-\sigma} \, \phi \, \psi^\dagger_{-\sigma} \psi_{-\sigma}
+ \frac{1}{2 e^2} (\d_y \phi)^2 + \frac{r}{2} \phi^2 \label{TwoPatch1}\eea 
with $e^2 = {2 m d^2}/{v_F}$, $r =  r_0 /(2 m d^2 )$, $\eta = 2 m$, and $\lambda_{s \sigma} = d_{s \sigma}/|d|$, and
we will henceforth drop the subscript $k_0$ on $L$. 
We note that as usual, the relation between the parameters of the effective theory and the original model should not be taken literally. Rather, in the critical regime, we have $r_0 - r_{0c} = Z_r (r-r_c)$, where $r_c$ and $r_{0c}$ denote the critical points of the effective theory and the microscopic theory respectively.
Moreover, the original fields and the fields defined in each patch of the Fermi surface are related by, 
\bea \phi(\vec{q},\omega) \sim Z_{\phi}^{1/2} K \phi_{patch}(K q_x, q_y, \omega), \quad \psi(\vec{q}, \omega) \sim Z_{\psi}^{1/2} K \psi_{patch}(K q_x, q_y, \omega) \label{relpatch}\eea
Note that the ``metric factors" $K$, $Z_r$, $e^2$, $Z_{\psi}$, $Z_{\phi}$ are generally dependent on the direction of the boson momentum $\hat{q}$ and the cut-off of the low-energy theory $\Lambda$.

For brevity, we will only present explicit calculations for the case that does not involve spin (Ising-nematic transition and spin-liquid); the extension of the results to the Ising ferromagnet case will be noted. Moreover, we extend the number of spin components (flavours) to $N$ from the physical value $N = 2$ with the view towards performing a large-$N$ expansion. For this purpose, it is convenient to rescale $e^2$ and $r$, yielding our Lagrangian in its final form
\beq L = \sum_{s=\pm} \psi^\dagger_{s} \Bigl(\eta \d_{\tau} - i s \d_x -  \d^2_y \Bigr) \psi_{s}   - \sum_{s = \pm} \lambda_s \, \phi \, \psi^\dagger_s \psi_s+ \frac{N}{2 e^2} (\d_y \phi)^2 + \frac{N r}{2} \phi^2 . \label{L} \eeq 
Here and below we suppress the flavour index. To reiterate, the Ising-nematic case has $\lambda_+ = \lambda_-$
and the spin-liquid case ({\em i.e.\/} Fermi surface coupled to U(1) gauge field) has $\lambda_+ = - \lambda_-$.

\section{One loop propagators}
\label{sec:oneloop}
To gain some insight into the low energy properties of the theory (\ref{L}), it is useful to compute the one loop boson and fermion self-energies. 
 


\begin{figure}[h]
\includegraphics*[width=1.8in]{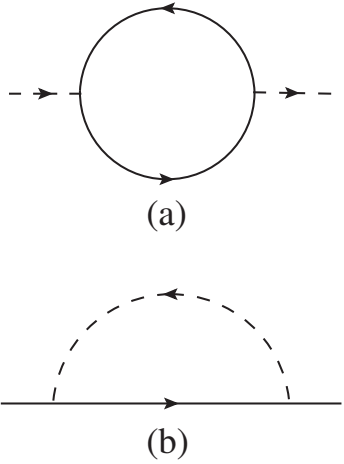}
\caption{One loop contributions to the (a) boson, and (a) fermion self-energies.}
\label{FigPi0}
\end{figure}

The one-loop boson polarization in Fig. \ref{FigPi0} a) is given by,
\beq \Pi_0(q) = N \int \frac{dl_\tau d^2 \vec{l}}{(2 \pi)^3} G^0_s(l) G^0_s(l+q)\eeq
We first evaluate this diagram with a bare
fermion propagator,
\beq G^0_s(k) = \frac{1}{-i \eta k_\tau + s k_x + k^2_y}\eeq
The resulting polarization function takes on a characteristic Landau-damped form,
\bea \Pi_0(q) &=& N \int \frac{ dl_\tau d l_y}{(2 \pi)^2} \frac{ i\left[\theta (l_\tau) - \theta(l_\tau + q_\tau) \right]  }{ 
- i \eta q_\tau + 2 q_y l_y + q_x + q_y^2} + (\vec{q} \to - \vec{q}) \nonumber \\
&=& \frac{N q_\tau}{2 \pi} \int \frac{d l_y}{2 \pi} \frac{ (-i) }{ 
- i \eta q_\tau + 2 q_y l_y + q_x + q_y^2} + (\vec{q} \to -\vec{q}) = 
c_b N \frac{|q_\tau|}{|q_y|}, \quad c_b = \frac{1}{4 \pi}. \label{Pi0}\eea
Note that $\eta$ has dropped out of the final result.
We are interested above only in the singular contribution to $\Pi_0$, and this is insensitive to orders
of integration: so unlike the conventional order, we have integrated over $l_x$ before $l_\tau$.
We include the RPA polarization bubble (\ref{Pi0}) into the bosonic propagator to obtain
\beq D(q) = \frac{1}{N} \left(c_b \frac{|q_\tau|}{|q_y|} + \frac{q^2_y}{e^2} + r \right)^{-1}. \label{propb}\eeq
Note that the $q_y^2$ term is not renormalized by the polarization contribution at this order,
and the bare co-efficient represents the phenomenological contribution of higher energy modes.

The one-loop correction to the fermion propagator is given by Fig. \ref{FigPi0} b). For simplicity, we work at the critical point and set $r = 0$. Then,
the fermion self-energy assumes a non-Fermi liquid form
\bea \Sigma_s(k) &=& - \int \frac{dl_\tau d^2 \vec{l}}{(2 \pi)^3} D(l) G^0_s(k-l) \nonumber \\
&=& - \frac{i}{2N}\int \frac{ d l_\tau d l_y}{(2 \pi)^2}  \left( c_b \frac{| l_\tau |}{| l_y |} + \frac{l_y^2}{e^2} \right)^{-1}
\times \mbox{sgn} (k_\tau - l_\tau ) \nonumber \\
&=& - \frac{i  c_f}{N} \mbox{sgn}(k_\tau) |k_\tau|^{2/3}, \quad c_f = \frac{2}{\sqrt{3}} \left(\frac{e^2}{4\pi}\right)^{2/3}. \label{Sigma0}\eea
Note, again, that $\eta$ has dropped out of the result.
Incorporating this correction into the fermion propagator,
\beq G_s(k) = \left(- \frac{i  c_f}{N} \mbox{sgn}(k_\tau) |k_\tau|^{2/3} + s k_x + k^2_y\right)^{-1} \label{propf}\eeq
Here we have dropped the bare fermion time derivative term proportional to $\eta$, which is irrelevant at low energies compared to the dynamically induced self-energy (\ref{Sigma0}).

As is well known,\cite{Polchinski} the one-loop expressions (\ref{Pi0}), (\ref{Sigma0}) actually satisfy the Eliashberg-like equations, in which the lines of Fig.~\ref{FigPi0} become self-consistent propagators. In what follows, we will use these self-consistent propagators (\ref{propb}), (\ref{propf}) in our calculations and drop self-energy corrections like those in Fig.~\ref{FigPi0}. 

\section{Scaling and Renormalization}
\label{sec:sr}

As has been argued by a numer of authors\cite{Polchinski,SSLee,sslee2,ALTSHULER}, a useful starting point for the renormalization group analysis of the theory (\ref{L}) is obtained by using the scaling,
\bea  k_x &\to& s^2 k_x,\,\, k_y \to s k_y,\,\, \omega \to s^3 \omega, \nonumber \\ 
\psi(x,y,\tau) &\to& s^2 \psi(s^2 x, s y, s^3 \tau),\,\, \phi(x,y,\tau) \to s^2 \phi(s^2 x, s y, s^3 \tau) \label{scaling}\eea
This scaling is suggested by the one-loop calculation of fermion and boson propagators in Eqs.~(\ref{propb}), (\ref{propf}).
The bare fermion time derivative term $\psi^\dagger \d_{\tau} \psi$ is irrelevant under this scaling, and so we will
take the limit $\eta \rightarrow 0^+$. Note that neither of the one loop corrections Eqs.~(\ref{Pi0}), (\ref{Sigma0}) depend
upon $\eta$. 

Alternatively, note that the scaling of time in (\ref{scaling}) could also have been derived by demanding that 
the `Yukawa coupling' $\lambda_s$ be invariant. This avoids the somewhat unnatural appeal to the one-loop
self-energy to set bare scaling dimensions, and yields all the scaling dimensions in (\ref{scaling}) by
a simple rescaling of the bare Lagrangian $L$ in Eq.~(\ref{L}). Of course, once we have set $\lambda_s$ to be invariant,
then the coupling $\eta$ becomes irrelevant. These features of the scaling analysis are shared by the theory
of the nematic transition in $d$-wave superconductors in Ref.~\onlinecite{huh}.

Note also the different scaling of spatial momenta $k_x$ and $k_y$ in Eq.~(\ref{scaling}). 
The main physical consequence of such momentum anisotropy is the effective decompactification of the 
Fermi surface, which allows one to focus on a theory with two Fermi patches.  
Also observe that under (\ref{scaling}) the $(\d_x \phi)^2$ part of the boson tree level action is irrelevant, which 
justifies omitting this term in eqs. (\ref{Ldim}), (\ref{L}). 
 
Apart from the fermion time derivative term and the relevant mass perturbation ($r \to s^{-2} r$), 
all the terms in the Lagrangian (\ref{L}) are marginal. 
Higher order perturbations to (\ref{L}), consistent 
with the $Z_2$ symmetry of the order parameter, such as a $\phi^4$ term, are irrelevant. 

We would like to note that for the case of the Ising-nematic (or $g$-wave) transition the low-energy action (\ref{L}) does not possess a $\phi \to - \phi$ symmetry. This is due to the fact that the direction of bosonic momentum $\vec{q}$ 
is transformed under $\pi/2$ rotations (reflections) and hence the physics is controlled by a different pair of patches of 
the Fermi surface. Hence, in principle, it is possible that in the kinematic regime of interest a $\phi^3$ term is generated by the renormalization group process. Such a term would be marginal under the scaling (\ref{scaling}). A linear term in $\phi$ can also be generated by the effective theory. However, the one-point function has momentum $\vec{q} = 0$ and, hence, does not belong to any particular kinematic regime. In practice, we can demand that the expectation value of $\phi$ is zero in the disordered phase by tuning the coefficient of the $\phi$-linear term. 
In any case, as we will show below, there exists a Ward identity, which guarantees that if these terms are initially zero, 
they are not generated by the RG of the low-energy theory (\ref{L}). Note that for the case of the spin-liquid or Ising ferromagnet transitions, the low energy theory (\ref{L}) respects the time reversal symmetry which maps Fermi patches at $k_0$ and $-k_0$ into each other and, hence, terms odd in $\phi$ are prohibited.

An important observation is that the theory (\ref{L}) lacks an expansion parameter. To see this, note that due to the rescaling performed in section \ref{sec:model}, the engineering dimensions, $[k_x] = [k_y]^2$, but the dimension of $\omega$ is kept independent. Then, the coupling constant $e^2$ has the dimensions $[k_y]^3/[\omega]$. Therefore, $e^2$ is a dimensionful quantity and cannot be used as an expansion parameter. Moreover, $e^2$ is actually the only parameter in the theory relating frequencies and momenta. Hence, its flow under RG is equivalent to an appearance of a non-trivial dynamical critical exponent.

Note that up to this point we have dropped an allowed relevant fermion chemical potential term,
\beq \Delta L = - \delta \, \psi^\dagger_s \psi_s \eeq
This term can be absorbed into the definition of the momentum $\vec{k}_0$ about which the theory is expanded and, thus,
is redundant (note, the scaling dimension $[\delta] = [k_x] = 2$). Nevertheless, it is convenient to leave this term in the Lagrangian 
for renormalization group purposes. We assume that when the theory is tuned to the criticality $r = r_c$ and the coefficient $\delta$ is set
to $\delta = \delta_c$, the Fermi surface passes through the points $\vec{k}_0, - \vec{k}_0$. 

We now discuss the renormalization of our theory. The Lagrangian contains four marginal operators, which each requires a renormalization constant. 
However, as we will argue below, emergent low-energy symmetries of the theory (\ref{L}) imply certain relations between these constants. Moreover, the two relevant operators, have the same bare dimension, $[r] = [\delta] = 2$. Thus, we need to consider possible mixing between these operators.

\subsection{Rotational Symmetry}
\label{sec:rot}

Observe that the initial shape of the Fermi surface does not enter the low-energy theory (\ref{L}). In fact, we could have started with a circular
Fermi surface with $k_F = m v_F$. This is reflected by the fact that Eq. (\ref{L}) has an emergent continuous ``rotational symmetry",
\beq \phi(x,y) \to \phi(x,y+\theta x), \quad \psi_s(x,y) \to e^{-i s(\frac{\theta}{2} y + \frac{\theta^2}{4} x)} \psi_s(x,y+\theta x) \eeq
Equivalently in momentum space,
\beq \phi(q_x,q_y) \to \phi(q_x - \theta q_y, q_y), \quad \psi_s(q_x, q_y) \to \psi_s \left(q_x - \theta q_y - s \frac{\theta^2}{4},  q_y + s \frac{\theta}{2}\right) \label{Rot}\eeq
Note that the rotation angle $\theta$ becomes non-compact and the rotation group becomes $\mathbb{R}$ instead of $U(1)$. This is a consequence of the effective decompactification of the Fermi surface. 
Moreover, due to the anisotropic scaling $\theta$ is now dimensionful $[\theta] = [k_y]$. In fact, the situation is analogous 
to the transformation of the Lorentz symmetry to Galilean invariance in the non-relativistic limit 
$\omega \ll c |\vec{q}|$. Here the role of $\omega$ is played by $q_x$ and the role of $|\vec{q}|$ by $q_y$. 

The symmetry (\ref{Rot}) implies the following form of the bosonic and fermionic Green's functions (we suppress the frequency dependence):
\bea D(q_x, q_y) &=& D(q_y) \label{slideb} \\ 
G_s(q_x, q_y) &=& G(s q_x + q^2_y). \label{slidef} \eea
In particular, the form of the fermionic Green's function implies that 
the terms $\psi^\dagger_s (- i s \d_x) \psi_s$ and $\psi_s (-\d^2_y) \psi_s$ in the Lagrangian (\ref{L}) must renormalize in 
the same way. Physically, this means that the curvature radius of the Fermi surface $K$ does not flow under 
RG ({\em i.e.\/} $K$ has a limit as the cutoff $\Lambda \to 0$).

\begin{figure}[t]
\includegraphics*[width=2in]{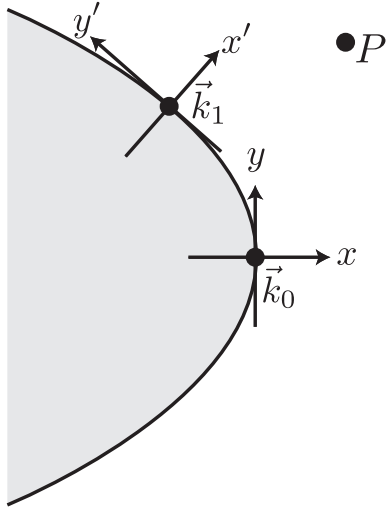}
\caption{The momentum of the fermion at point $P$ can be measured with respect to either the co-ordinate
system at $\vec{k}_0$, or that at $\vec{k}_1$.}
\label{fig:shift}
\end{figure}

The identities (\ref{slideb},\ref{slidef}) ensure that the Green's functions at a given physical momentum
remain invariant under small changes
in the choice of the points $\pm \vec{k}_0$ on the Fermi surface about which the field theory is defined.
Let us demonstrate this explicitly using Fig.~\ref{fig:shift}.
We set the co-ordinate system so that $\vec{k}_0 = (0,0)$, and measure the momentum of a fermion
at the point $P$ to be $(q_x, q_y)$. Now let us shift to the field theory defined at the Fermi surface
point $\vec{k}_1 = (\kappa_x, \kappa_y)$. As this point has to be on the Fermi surface, we have
$\kappa_x + \kappa_y^2 = 0$. We denote the co-ordinates of the point $P$ in the new co-ordinate
system by $(q_x', q_y')$. These are obtained from the old co-ordinates by a shift in origin followed by a rotation
by an angle $\theta$, where $\tan \theta = 2 \kappa_y$; this yields 
\begin{eqnarray}
q_x' &=& q_x - \kappa_x + 2 \kappa_y ( q_y - \kappa_y) \nonumber \\
q_y' &=& q_y - \kappa_y~~~~,
\label{eqshift}
\end{eqnarray}
where we only keep terms to the needed accuracy of $\mathcal{O} (x, y^2)$. It can now be verified that
$ q_x' + q_y'^2 = q_x + q_y^2 $, and so by Eq.~(\ref{slidef}) the fermion Green's function remains invariant
under the change in the Fermi surface reference point. Also, by choosing $\kappa_y = q_y$ we can set
$q_y'=0$, and then $q_x + q_y^2$ is identified as the invariant measuring the 
distance between $P$ and the closest
point on the Fermi surface.
For the boson Green's function, there is no shift in origin
of the co-ordinates, and the corresponding transformation is $q_x' = q_x + 2 \kappa_y q_y$, $q_y' = q_y$,
and this remains invariant under Eq.~(\ref{slideb}).

These invariances are essential in ensuring the consistency of our description of each pair of time-reversed
Fermi surface points by a separate 2+1 dimensional field theory. Note that such a consistency requirement
would not have arisen if we had used a 1+1 dimensional field theory at each Fermi surface point,\cite{Luther,Marston,KwonMarston,haldane,NetoFradkin,LF07,LFBFO06}  because
then every fermion momentum would appear only in the theory defined at the closest point on the Fermi surface. In our case, we are free to use the 2+1 dimensional theory at this closest point,
or at any of the neighboring points.

Before concluding this section, we would like to point out that in the case of the Ising-nematic transition, the ``rotational symmetry" (\ref{Rot}) is not related in any way to ``large" rotations by $\pi/2$, which are actually not implemented in the low-energy theory.

\subsection{Ward Identities}

We now examine the consequences of Ward identities associated with the global symmetries of Eq.~(\ref{L}). 
Similar consequences were implicit in the analysis of the superconducting case in Ref.~\onlinecite{huh}.
Here we will present a more formal analysis, which also shows that Eq.~(3.20) in Ref.~\onlinecite{huh} holds
to all orders in $1/N$.

The low energy theory (\ref{L}) has two continuous global $U(1)$ symmetries. The first of these is related to the conservation of particle number,
\beq U(1)_F: \psi_+ \to e^{i \alpha} \psi_+, \quad \psi_- \to e^{i \alpha} \psi_- \eeq
The conserved current associated with this symmetry is,
\beq (j_\tau, j_x, j_y)_F = (i \eta (\psi^\dagger_+ \psi_+ + \psi^\dagger_- \psi_-), \psi^\dagger_+ \psi_+ - \psi^\dagger_- \psi_-, -i (\psi^\dagger_+ \overleftrightarrow{\d}_y \psi_+ + \psi^\dagger_- \overleftrightarrow{\d}_y \psi_-))\label{jF}\eeq
For the spin-liquid problem, the gauge field $\phi$ couples precisely to the $x$ component of $j_F$.

The second $U(1)$ symmetry is lattice translation. Indeed, $\psi_+$ and $\psi_-$ come from opposite points in the Brilloin zone and, hence, transform under general lattice translations as,
\beq U(1)_T: \psi_+ \to e^{i  \alpha} \psi_+,\quad \psi_- \to e^{-i  \alpha} \psi_- \eeq
The conserved current associated with this symmetry is 
 \beq (j_\tau, j_x, j_y)_T = (i \eta (\psi^\dagger_+ \psi_+ - \psi^\dagger_- \psi_-), \psi^\dagger_+ \psi_+ + \psi^\dagger_- \psi_-, -i (\psi^\dagger_+ \overleftrightarrow{\d}_y \psi_+ - \psi^\dagger_- \overleftrightarrow{\d}_y \psi_-))\label{jT}\eeq
Observe that the Ising-nematic order parameter $\phi$ couples to the $x$ component of $j_T$. Note that despite the similarity of the spin-liquid and Ising-nematic problems, there is an important difference. In the spin-liquid case, the gauge field couples to the fermion current on all energy scales. In the case of the Ising-nematic transition, the order parameter couples to a conserved current only at low energies.

We note in passing that for an Ising ferromagnet transition, the current to which the order parameter couples is related to the symmetry,
\beq U(1)_I: \psi_{+\uparrow} \to e^{i  \alpha} \psi_+,\quad \psi_{-\uparrow} \to e^{-i  \alpha} \psi_-,\quad \psi_{+\downarrow} \to e^{- i  \alpha} \psi_{+\downarrow},\quad \psi_{-\downarrow} \to e^{i  \alpha} \psi_{-\downarrow} \eeq
In fact, this is not a symmetry of the underlying theory, but only of the low-energy Lagrangian (\ref{Ldim}). The symmetry is broken by four-Fermi interactions, which are however irrelevant under (\ref{scaling}). 

Current conservation implies that the insertion of $\d_{\tau} j_{\tau} + \d_x j_x + \d_y j_y$ into any correlation function is zero, up to contact terms (we have dropped the current subscript; the current, which couples to the order parameter is implicitely assumed). We note that the temporal component of the currents (\ref{jF}), (\ref{jT}) has a coefficient $\eta$ in front and, therefore, can be set to zero in the kinematic regime of interest. We, thus, have $\d_x j_x + \d_y j_y \sim 0$. Defining the one-particle irreducible polarization function,
\beq \Pi_{ij}(q) = \int d\tau d^2 x e^{i q_\tau \tau - i \vec{q} \cdot \vec{x}} \langle j_i(x) j_j(0)\rangle_{1PI}\eeq
we have
\beq q_x \Pi_{xx}(q) + q_y \Pi_{y x}(q) = 0 \label{WI1}\eeq
We note that $\Pi_{xx}(q) = \Pi_{xx}(q_\tau,q_y)$ is precisely the irreducible boson self-energy. Hence,
\beq \Pi_{yx}(q_\tau,q_x,q_y) = - \frac{q_x}{q_y} \Pi_{xx}(q_\tau,q_y)\nn\eeq
Power counting indicates that $\Pi_{xx}$ has the following UV structure
\beq \Pi_{xx}(q_\tau,q_y) \stackrel{UV}{=} K_1  + K_2 r + K_3 q^2_y\eeq
where $K_1 \sim \Lambda^2$, $K_2, K_3 \sim \log \Lambda$ and $\Lambda$ is the $UV$ cut-off with dimensions of $q_y$. For $\Pi_{yx}(q_\tau,q_x, q_y)$ 
to have an analytic $UV$ behaviour (as again expected from power counting), we must have
\beq K_1 = K_2 = 0 \nn \eeq
Thus, the coefficient of the mass operator $\phi^2$ requires no renormalization ({\em i.e.\/} the metric factor $Z_r$ has a limit as $\Lambda \to 0$). 

An interesting question is whether the polarization function $\Pi_{xx}$ actually vanishes for $q_y \to 0$ as suggested by Eq. (\ref{WI1}). However, for finite $q_\tau$ we already know from one-loop calculations that such a limit does not exist within the scaling regime, as
\beq \Pi_{xx}(q_\tau, q_y)_{1loop} = c_b \frac{|q_\tau|}{|q_y|}, \quad \Pi_{yx}(q_\tau, q_x,q_y)_{1loop} = -c_b\frac{q_x}{q_y} \frac{|q_\tau|}{|q_y|}\nn\eeq
However, one might hope that the limits $\lim_{q_y \to 0} \lim_{q_\tau \to 0} \Pi_{xx}(q_\tau,q_y), \Pi_{xy}(q_\tau,q_x,q_y)$ do exist. In this case, we would conclude,
\beq  \lim_{q_y \to 0} \lim_{q_\tau \to 0} \Pi_{xx}(q_\tau,q_y) = 0 \label{Pinonr}\eeq
which would be a stronger statement than the non-renormalization of the mass term. Otherwise, if the limit above exists only for $\Pi_{xx}$ by not $\Pi_{xy}$ then,
\beq \lim_{q_y \to 0} \lim_{q_\tau \to 0} \Pi_{xx}(q_\tau,q_y) = c_r r \eeq
with $c_r$ - some universal constant. We have explicitly checked that to three loop order $c_r = 0$ and the strong form of the non-renormalization identity Eq. (\ref{Pinonr}) holds. 

One can generalize the discussion above to higher order correlation functions of the order parameter. Ward-identities imply that the effective potential for the $\phi$ field is not renormalized from its tree-level form,
\beq V(\phi) = \frac{r}{2} \phi^2 \label{Vnonren} \eeq
This property is also shared by the theory of the nematic transition in a $d$-wave superconductor.\cite{kim,huh} In particular, no $\phi^3$ term is induced in the Lagrangian by the renormalization group process if this term is originally zero. (Note that if 
a $\phi^3$ term is initially present, correlation functions of currents no longer coincide with the correlation functions of the order parameter, and the Ward identities do not constrain the renormalization properties of the theory). The effective potential (\ref{Vnonren}) becomes unstable for $r < 0$. Thus, we expect that in the ordered phase the theory is controlled by dangerously irrelevant operators, such as $\phi^4$. 

Finally, one can derive a Ward identity for the fermion boson vertex,
\beq q_x \Gamma_x(q,p,p+q) + q_y \Gamma_y(q, p, p+q) = G^{-1}(p+q) - G^{-1}(p) \eeq
with
\beq \Gamma_i(q,p,p+q) = \int d x_\tau d^2 x  d y_\tau d^2y  e^{-i q_\tau x_\tau + i \vec{q} \cdot \vec{x}} e^{i (p+q)_\tau y_\tau - i (\vec{p} + \vec{q}) \cdot \vec{y}} \langle j_i(x) \psi(y) \psi^\dagger(0)\rangle_{1PI} \eeq 
\beq G(p) = \int d\tau d^2 x e^{i q_\tau \tau - i \vec{q} \cdot \vec{x}} \langle \psi(x) \psi^{\dagger}(0)\rangle \eeq
$\Gamma_x$ is precisely the irreducible fermion-boson vertex. Power counting gives $UV$ structure of $\Gamma_x$ and $G^{-1}$ as,
\bea &&\Gamma_x(q,p,p+q) = C_1\\
&&G^{-1}(p) = C_2 + C_3(p_x + p^2_y) \eea
Thus, for the $UV$ behaviour of $\Gamma_y$ to be analytic in external momenta, $C_1 = C_3$. Therefore, the vertex and the fermion self-energy renormalize in the same way. Hence, the boson field requires no field-strength renormalization ({\em i.e.\/} the metric factor $Z_{\phi}$ has a limit as $\Lambda \to 0$).

Before concluding this section, we would like to note that perturbation theory based on self-consistent propagators (\ref{propb}), (\ref{propf}) actually does not respect the Ward identities. This is due to the fact that these one-loop propagators include the fermion self-energy correction, but not the vertex correction. However, since the fermion self-energy is only frequency dependent, Ward identities involving currents at zero external frequency are still respected.

\subsection{RG equations}
From the discussion above, we conclude that at criticality, our theory needs only two renormalizations: a rescaling of the field strength of the fermion field $\psi$ and a renormalization of $e^2$,
\beq \psi = Z^{1/2}_\psi \psi_r, \quad e^2 = Z_e e^2_r \label{defZ} \eeq
Here the subscript $r$ denotes renormalized quantities and we define renormalized irreducible correlation functions of $n_b$ boson and $n_f$ fermion fields as,
\beq \Gamma^{n_b,n_f}_r = Z^{n_f/2}_\psi \Gamma^{n_b,n_f} \label{Gammardef}\eeq 
Both $Z_\psi$ and $Z_e$ are functions of $\Lambda/\mu$ where $\mu$ is a renormalization scale (which we choose to have dimensions of $q_y$) and of the number of fermion flavours $N$. As $e^2$ is dimensionful, $Z_\psi$ and $Z_e$ cannot depend on it. We introduce the anomalous dimensions,
\bea b &=&   \Lambda \frac{\d}{\d \Lambda} \log Z_e \label{betae}\\
\eta_\psi &=&  - \Lambda \frac{\d}{\d \Lambda} \log Z_\psi \label{betaf} \eea
The constants $\eta_\psi$ and $b$ are expected to be pure universal numbers, independent of $\Lambda/\mu$. 

Away from criticality, we recall that by the Ward identity, the coupling $r$ does not renormalize. On the other hand, the coupling $\delta$ can pick up 
a renormalization linear in $r$,
\beq \delta = \delta_c + \delta_r + Z_{r \delta} e^2_r r \label{deltaren}\eeq
with $Z_{r \delta}$ again a function of $\Lambda/\mu$ only. In what follows, we denote $\delta - \delta_c$ as $\delta$ for brevity. Note that there is no renormalization constant in front of $\delta_r$ since a finite change in $\delta$ only shifts the value of $k_x$ in correlation functions:
\beq \Gamma^{n_b,n_f}(\{p\}, \delta+a) =  \Gamma^{n_b,n_f}(\{p -s a \hat{x}\}\,, \delta) \label{shiftdelta}\eeq
where $s = \pm 1$ for momenta of fermions $\psi_{\pm}$ and $s = 0$ for boson momenta. We let,
\beq \alpha = Z_e^{-1} \Lambda \frac{\d}{\d \Lambda} Z_{r \delta} \label{alphadef} \eeq

Now, differentiating Eq. (\ref{Gammardef}) we obtain the renormalization group equations
\beq \left(\Lambda \frac{\d}{\d \Lambda} + b e^2 \frac{\d}{\d e^2} + \alpha e^2 r \frac{\d}{\d \delta} - \frac{n_f}{2} \eta_\psi\right) \Gamma^{n_b, n_f}(\{p_y\},\{p_x\},\{\omega\}, r, \delta, e^2, \Lambda) = 0 \label{RG1}\eeq
It is convenient to get rid of the derivative with respect to $\delta$ in Eq. (\ref{RG1}). To do so, let the location of the Fermi-surface of fermion $\psi_+$ at finite $\delta$ and $r$ be given by $k_x + k^2_y = \Delta k(r,\delta, e^2, \Lambda)$. Then, $\Delta k$ is clearly a physical quantity and must satisfy,
\beq  \left (\Lambda \frac{\d}{\d \Lambda} + b e^2 \frac{\d}{\d e^2} + \alpha e^2 r \frac{\d}{\d \delta}\right) \Delta k(r,\delta, e^2, \Lambda) = 0. \label{RGK}\eeq
We will solve this equation shortly. However, first note that 
\beq \frac{\d \Delta k}{\d \delta} = 1. \label{dK}\eeq
 Now, it is convenient to expand momenta around the physical Fermi-surface, defining,
\beq \tilde{\Gamma}^{n_b,n_f}(\{p\},r, \delta, e^2, \Lambda) = \Gamma^{n_b,n_f}(\{p+s \Delta k(r, \delta, e^2, \Lambda) \hat{x} \},r, \delta, e^2, \Lambda)\eeq
The resulting $\tilde{\Gamma}$ is independent of $\delta$ and by Eqs. (\ref{shiftdelta}), (\ref{RG1}), (\ref{RGK}), (\ref{dK}) satisfies,
\beq \left(\Lambda \frac{\d}{\d \Lambda} + b e^2 \frac{\d}{\d e^2}  - \frac{n_f}{2} \eta_\psi\right) \tilde{\Gamma}^{n_b, n_f}(\{p_y\},\{p_x\},\{\omega\}, r,  e^2, \Lambda) = 0\eeq

By dimensional analysis,
\beq \tilde{\Gamma}^{n_b, n_f} = \Lambda^{6 - 2 n_f - 2 n_b} (e^2)^{n_f/2 - 1} f^{n_b, n_f}\left(
\bigg\{\frac{p_y}{\Lambda}\bigg\},\bigg\{\frac{p_x}{\Lambda^2}\bigg\},\bigg\{\frac{\omega e^2}{\Lambda^3}\bigg\},\frac{\Lambda^2 r}{\mu^2}
\right)
\eeq 
and solving the RG equation, we obtain
\beq f^{n_b, n_f}(s\{\tilde{p}_y\}, s^2 \{\tilde{p}_x\}, s^{3-b} \{\tilde{\omega}\}, s^{2-b} \tilde{r}) = s^{6 - b + (b - \eta_\psi - 4) n_f/2 - 2 n_b}
f^{n_b, n_f}(\{\tilde{p}_y\}, \{\tilde{p}_x\}, \{\tilde{\omega}\}, \tilde{r})\eeq
Hence, the critical theory is invariant under,
\beq p_y \to s p_y, \quad p_x \to s^2 p_x, \quad \omega \to s^z \omega \label{pxpyo} \eeq
with 
\beq z = 3 - b ,\label{3mb} \eeq
where $z$ is the dynamic critical exponent. Note that we have defined $z$ with reference to length scales associated with
directions tangent to the Fermi surface ($y$); as indicated in (\ref{pxpyo}), length scales orthogonal to the Fermi 
surface scale as the square of length scales tangent to the Fermi surface. 
Moreover, if we define $\xi$ as the correlation length along the $y$ direction then upon approaching the critical point, $\xi \sim r^{-\nu}$, with
\beq \nu = \frac{1}{z-1}. \label{nuz} \eeq

Note that by combining Eqs.~(\ref{defZ},\ref{betae},\ref{3mb}) we can write down the RG equation for the
coupling $e$:
\beq \Lambda \frac{\partial e^2}{\partial \Lambda}\bigg|_{e^2_r,\mu} = -(z-3) e^2 .
\eeq
This shows that the renormalization of the coupling $e$ is directly related to the dynamic critical exponent,
as we had claimed earlier.

Now, let us consider a few explicit examples of correlation functions. For the bosonic two-point function we have,
\beq D^{-1}(q_y, \omega) = r g \left(q_y (r e^2 \Lambda^{z-3})^{-\frac{1}{z-1}}, \omega (r^z e^2 \Lambda^{z-3})^{-\frac{1}{z-1}}
\right)\eeq
Note that, 
\beq \lim_{q_y \to 0} \lim_{\omega \to 0} D^{-1}(q_y, \omega) = r g(0,0) \eeq
{\em i.e.\/} the Ising-nematic susceptibility satisfies $\chi \sim r^{-\gamma}$ with the exponent
\beq \gamma = 1 . \eeq
We may also write more succinctly,
\beq D^{-1}(q_y, \omega) \propto \xi^{-(z-1)} g(q_y \xi, \omega e^2 \Lambda^{z-3} \xi^z) \label{Dscal} \eeq

So far, we have been concentrating on a fixed direction of bosonic momentum $\vec{q}$. Now let us study the dependence of the result on $\hat{q}$. Using Eq. (\ref{relpatch})
\beq D^{-1}(\vec{q}, \omega) = Z^{-1}_{\phi} K^{-1} Z^{-1}_r r_0 g \left(|\vec{q}| (Z^{-1}_r e^2 \Lambda^{z-3} r_0)^{-\frac{1}{z-1}}, \omega (Z^{-z}_r e^2 \Lambda^{z-3} r^z_0)^{-\frac{1}{z-1}} \right) \eeq
where for brevity $r_0$ is taken to denote the deviation from the critical point. We concentrate on the static limit $\omega = 0$. In a Fermi liquid, the susceptibility must have a continuous limit as $\vec{q} \to 0$. Therefore, we conclude that the combination $Z_{\phi} K Z_r$ must be independent of the direction $\hat{q}$. This is quite plausible, as neither of the constants run under RG. 

Now let us look at the behaviour of susceptibility at the critical point,
\beq D^{-1}(q_y, \omega) = \frac{q^{z-1}_y}{e^2 \Lambda^{z-3}} h\left(\frac{\omega e^2 \Lambda^{z-3}}{q^z_y}\right) \label{suscrit}\eeq
In particular, the static susceptibiltiy satisfies,
\beq D^{-1}(\vec{q}, 0) \sim a(\hat{q}) |\vec{q}|^{z-1} \eeq 

In the context of the spin-liquid problem, many studies \cite{KIM94,ybkim2,stern1,stern2,read,rshankar} 
examined the structure of the higher loop corrections to the susceptibility.
In particular, Kim {\em et al.} \cite{KIM94} examined two-loop corrections
to $\mbox{Im} \, D^{-1} (\vec{q}, \omega)$ for real frequencies $|\omega| \ll |\vec{q}|$, and found no corrections to the leading answer
$\sim \omega/|q_y|$ in Eq.~(\ref{propb}); Fermi liquid arguments were 
made  \cite{KIM94,ybkim2,stern1,stern2,read}  
that this functional form held at
higher orders. However, this result by itself does not fix the value of $z$; indeed, $\mbox{Im} \, D^{-1} (\vec{q}, \omega) \sim \omega/|q_y|$ is consistent with the scaling form (\ref{suscrit}) for {\em any\/} $z$.
These studies also implicitly assumed a Fermi liquid picture with $D^{-1} (\vec{q}, \omega = 0) \sim \vec{q}^2$, and this
does imply $z=3$. We will examine $D^{-1} (\vec{q}, \omega = 0)$ up to 3 loops in 
Section~\ref{sec:alpc}, and find no
correction to $z=3$.


Proceeding to the fermion Green's function,
\beq G^{-1}_s(\vec{k}, \omega) = \Lambda^2 \left(\frac{r e^2}{\Lambda^2}\right)^{\frac{2-\eta_\psi}{z-1}} L \left(k (r e^2 \Lambda^{z-3})^{-\frac{2}{z-1}}, \omega (r^z e^2 \Lambda^{z-3})^{-\frac{1}{z-1}}\right)
\eeq
with $k = s k_x + k_y^2$ - the distance to the Fermi surface. More compactly,
\beq G^{-1}(\vec{k}, \omega) \propto \xi^{-(2-\eta_\psi)} L\left(k \xi^2, \omega e^2 \Lambda^{z-3} \xi^z \right) \label{Gscal}\eeq
A crucial property of the theory that is manifested by the above expression is that the ``fermionic correlation length" scales as the square of the ``bosonic correlation length".

For $\omega \ll \xi^{-z}$, $k \ll \xi^{-2}$ we expect the fermion Green's function to assume a Fermi-liquid form,
\beq G(\vec{k}, \omega) = \frac{Z}{-i \omega + v_F k} \eeq
By matching to the scaling form,
\beq v_F \sim \xi^{-(z-2)}, \quad Z \sim \xi^{-(z+\eta_\psi - 2)}\eeq
Notice that both the Fermi velocity $v_F$ and the residue $Z$ tend to zero as we approach the critical point, albeit with different power laws. Finally, at the quantum critical point,
\beq G^{-1}(\vec{k}, \omega) = \Lambda^{\eta_\psi} k^{1-\eta_\psi/2} P\left(\frac{\omega e^2 \Lambda^{z-3}}{k^{z/2}}\right),\label{Gcrit}
\eeq
where we reiterate that $k = s k_x + k_y^2$ is the distance to the Fermi surface.
In particular, the self-energy on the Fermi surface scales as,
\beq G^{-1}(0, \omega) \sim \omega^{(2-\eta_\psi)/z}\eeq
and the static self energy,
\beq G^{-1}(\vec{k},0)  \sim k^{1-\eta_\psi/2} \eeq
Moreover, from Eq. (\ref{Gcrit}) we can obtain the tunneling density of states,
\beq N(\omega) = \int \frac{d^2 k}{(2\pi)^2} A(\vec{k}, \omega) \label{DOS1}\eeq
where
\beq A(\vec{k}, \omega) = -\frac{1}{\pi} \mathrm{Im} G(\vec{k}, i \omega \to \omega + i 0^+)\eeq
The $\vec{k}$ integral in Eq. (\ref{DOS1}) factorizes into integrals over components along and perpendicular
to the Fermi surface. The former gives a factor proportional to the perimeter of the Fermi surface, while the 
later yields the frequency dependence,
\beq N(\omega) \sim \omega^{\eta_\psi/z} \label{DOS2}\eeq
We remind the reader that the expression in Eq. (\ref{DOS2}) corresponds to the physically observable electron tunneling density of states only in the case of a nematic transition, as for the spin/charge-liquid problem, the physical electron operator is a product of $\psi$ and a boson operator. 

Related scaling forms for the fermion Green's function were discussed on a phenomenological 
basis by Senthil.\cite{senthil} However his definition of $z$ differs from ours. We define it using the fermion
momentum parallel to the Fermi surface, because this is the natural momentum scale appearing
also in the boson correlations. He defines it 
by the fermion momentum orthogonal to the Fermi surface, which scales as the square of the 
parallel momentum.

Finally, let us discuss the shift of the Fermi surface $\Delta k$. Using Eq. (\ref{dK}) in the RG equation (\ref{RGK}),
we obtain,
\beq \Delta k = \frac{\alpha}{z-3} r e^2 + C_k (r e^2 \Lambda^{z-3})^{2 \nu} + \delta  \label{dKsol1}\eeq
Thus, the shift of the Fermi surface upon deviation from the critical point receives two contributions: one analytic in $r$ and the other non-analytic. Reexpressing the second contribution in terms of the correlation length, 
\beq \Delta k =  \frac{\alpha}{z-3} r e^2 + \tilde{C}_k \xi^{-2} + \delta \label{dKsol2}\eeq
where the coefficient $\tilde{C}_k$ is expected to be universal. We would like to point out that the case $z = 3$ has to be treated separately. In this situation one obtains,
\beq \Delta k = \frac{\alpha r e^2}{2} \log \frac{r e^2}{\Lambda^2}  = - \hat{C}_k  \xi^{-2} \log (\Lambda \xi) + \delta \label{dKsollog}\eeq
with $\hat{C}_k$ again universal.

The value of the Fermi surface shift $\Delta k$ can be used to compute the compressibility, $\frac{\d n}{\d \mu}$, where $\mu$ is the physical chemical potential. Indeed, by Luttinger's theorem the change in density can be obtained as,
\beq \delta n = \frac{N}{(2 \pi)^2} \int d s \, \Delta k (\theta) \eeq
where the integral is over the circumference of the Fermi-surface. The main question is how does the chemical potential enter our low-energy theory. If $\mu$ only couples to the operator $\psi^{\dagger} \psi$, renormalizing the value of $\delta$, then from  Eqs. (\ref{dKsol1}), (\ref{dKsollog}) we would conclude that the compressibility tends to a constant and has no interesting corrections near the quantum critical point. On the other hand, if the coupling $r$ has a non-trivial $\mu$ dependence, then we would conclude,
\beq \frac{\d n}{\d \mu} = \kappa_0 + \kappa_1 \xi^{z - 3}, \quad z \neq 3 \label{comprz}\eeq
\beq \frac{\d n}{\d \mu} = \kappa_0 + \hat{\kappa}_1 \log \Lambda \xi,\quad  z = 3 \label{compr3}\eeq
Note that for $z \ge 3$ the above forms imply that the compressibility diverges as we approach the critical point. 
\section{Anomalous exponents to three loops}
\label{sec:aetl}

In this section, we evaluate the exponents $z$ and $\eta_\psi$ to three loop order. We find that the exponent $\eta_\psi$ is non-zero at this order. The value of $\eta_\psi$ is not suppressed in the large-$N$ limit. On the other hand, the dynamical critical exponent $z$ remains unrenormalized from its RPA value $z = 3$ to this order. Moreover, 
in the large-$N$ limit, the boson self energy acquires a finite correction of order $N^{3/2}$, which is larger than the bare value (order $N$). Finally, we find that the constant $\alpha$ in Eq. (\ref{alphadef})  associated with the shift of the Fermi-surface away from criticality is non-zero at three loop order. We note that the $N^{3/2}$ correction to the boson self-energy and the non-zero $\eta_\psi$ are only present for the Ising-nematic and spin-liquid universality classes, and are absent for the Ising ferromagnet transition. 

\subsection{Dynamical critical exponent}
\label{sec:alpc}

\begin{figure}[h]
\begin{center}
\includegraphics*[width=2.5in]{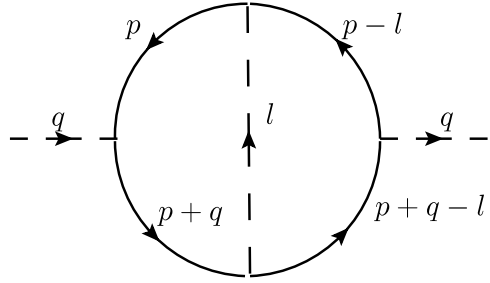}
\caption{Two loop corrections to the polarization.}
\label{PiTwoLoop}
\end{center}
\end{figure}

 Let us first address the question of renormalization of $e^2$. At two-loops the only correction to the static boson-self energy $\Pi(q_\tau = 0, \vec{q})$, which is not already taken into account by the solution to self-consistent Eliashberg equations is given in Fig. \ref{PiTwoLoop}. However, this diagram vanishes when the external frequency is equal to zero. Indeed, as pointed out in
Ref.~\onlinecite{SSLee}, any diagram with fermions from a single patch, in which the fermion propagators involve a sum of two or less internal momenta, vanishes in the static limit (one picks the internal frequency with the largest absolute value and integrates over the corresponding $x$ component of the momentum. All poles will be in the same half-plane). Actually, a calculation presented in Appendix \ref{app:comp} shows that the diagram in Fig. \ref{PiTwoLoop} vanishes for any external frequencies and momenta.

\begin{figure}[h]
\begin{center}
\includegraphics*[width=4in]{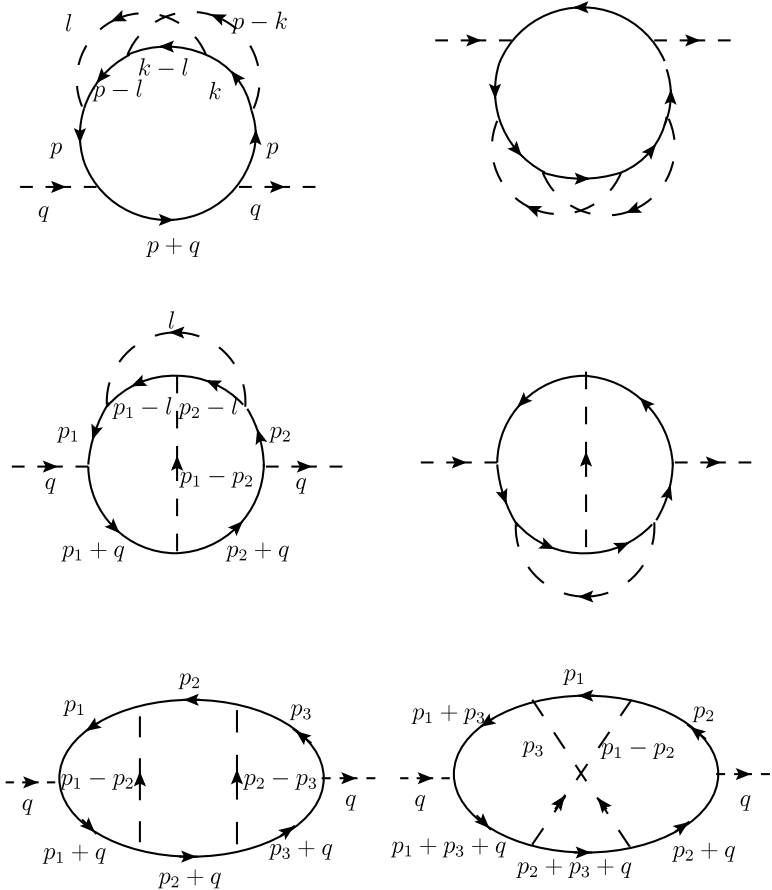}
\caption{Three loop corrections to the boson self-energy with one fermion loop.}
\label{PiThreeLoop}
\end{center}
\end{figure}

\begin{figure}[h]
\begin{center}
\includegraphics*[width=2.5in]{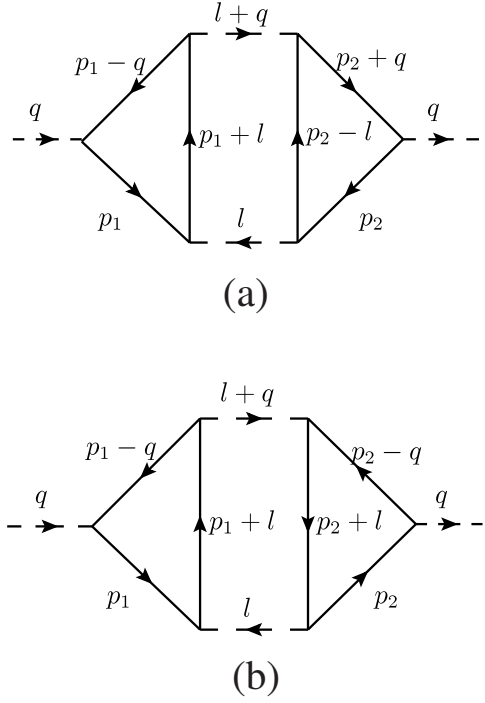}
\caption{Aslamazov-Larkin type three loop contributions to the boson self-energy.}
\label{FigAzLark}
\end{center}
\end{figure}

The three loop corrections to $\Pi(q)$ are shown in Figs. \ref{PiThreeLoop} and \ref{FigAzLark}. By the argument described above, all of these diagrams vanish when the external frequency is zero if all the fermions are from the same patch. Hence, the only non-zero corrections to $\Pi(q_\tau = 0, \vec{q})$ come from the Aslamazov-Larkin type diagrams, Fig. \ref{FigAzLark},
\begin{figure}
\begin{center}
\includegraphics*[width=2.5in]{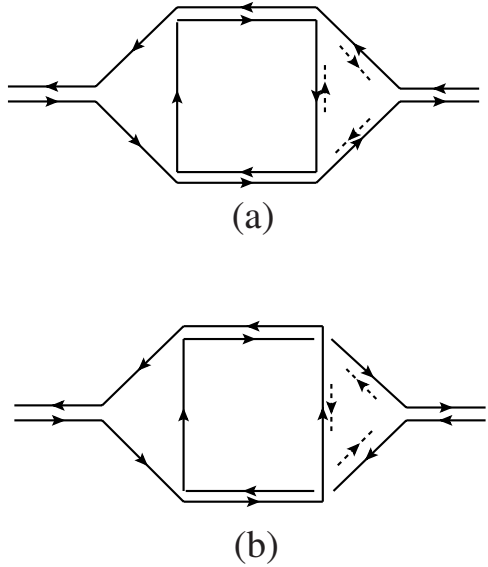}
\end{center}
\caption{Double line representation of Ref.~\onlinecite{SSLee} applied to the Aslamazov-Larkin diagrams in Fig. \ref{FigAzLark}. The fermions in the two loops
are assumed to come from opposite patches. We have reversed the directions of the fermion propagators
from the second patch, and the dotted arrows indicate the true directions of the fermion momenta. }\label{FigAzLarkDouble}
\end{figure}
\beq \delta^3 \Pi(q) = - \frac{1}{2} \int\frac{dl_\tau d^2 \vec{l}}{(2 \pi)^3} \Gamma^3(q,l,-(l+q)) \Gamma^3(-q,-l,l+q) D(l) D(l+q)\eeq
Here $\Gamma^3$ is the fermion-induced cubic boson vertex, which receives contribution from the two fermion patches,
\beq \Gamma^3 = \Gamma^3_+ + \Gamma^3_-\eeq
\beq \Gamma^3_s(l_1,l_2,l_3) = N \lambda^3_s (f_s(l_1,l_2,l_3) + f_s(l_2,l_1,l_3))\eeq
\beq f_s(l_1,l_2, l_3) = \int \frac{dp_\tau d^2 \vec{p}}{(2\pi)^3} G_s(p) G_s(p-l_1)G_s(p+l_2)\label{f}\eeq
The diagrams where the fermions in the two loops come from the same patch give a vanishing contribution to $\Pi(q_{\tau} = 0, \vec{q})$. Thus, to three loops,
\bea \delta^3 \Pi(q_{\tau} = 0, \vec{q}) &=& - \frac{1}{2} \int\frac{dl_\tau d^2 \vec{l}}{(2 \pi)^3} \Gamma^3_+(q,l,-(l+q)) \Gamma^3_-(-q,-l,l+q) D(l) D(l+q) + (q \to - q)\nn\\
&=& - \lambda^3_+ \lambda^3_- N^2 \int \frac{dl_\tau d^2 \vec{l}}{(2 \pi)^3} \Bigl[ f_+(q,l,-(l+q)) (f_-(-q,-l, l+q) 
\nn \\
&~&~~~~~~~~~+ f_-(-q,l+q,-l)) D(l) D(l+q) \Bigr]
+ (q\to -q).  \label{deltaPi} \eea
The two terms in brackets in the equation above originate respectively from diagrams in Figs. \ref{FigAzLark} a) and b). Converting these diagrams into the double line representation of Ref.~\onlinecite{SSLee}, we obtain Figs. \ref{FigAzLarkDouble} a) and b). [We remark that the genus expansion of Ref.~\onlinecite{SSLee} was developed for a theory with only a single Fermi-surface patch. The extension to the present case of a pair of time reversed patches is simple: a reversal of the direction of loops with fermions from the second patch reduces the problem to that with one patch only. The diagrams in Fig. \ref{FigAzLarkDouble} have their lines reversed precisely in this way. The additional dotted arrow besides each propagator indicates the true direction of fermion momentum.] In this representation, the graph a) contains a loop  while the graph b) does not. As a result, in the genus expansion of Ref.~\onlinecite{SSLee}, the diagram in Fig. \ref{FigAzLark} a) is enhanced to $O(N)$, while the diagram in Fig. \ref{FigAzLark} b) is of $O(1)$. However, we will see that the diagrams are actually individually ultra-violet divergent, as a result the counting of Ref.~\onlinecite{SSLee} 
is inapplicable here. It turns out that the sum of the diagrams is $UV$ finite and of $O(N^{3/2})$. 

We give details of the evaluation of Eq.~(\ref{deltaPi}) in Appendix~\ref{app:comp}, where we find
\beq \delta^3 \Pi(q_\tau=0, \vec{q}) = C \lambda_+ \lambda_- \frac{q^2_y}{e^2}
\label{deltaPires}
\eeq
In the large-$N$ limit, the coefficient $C$ is given by,
\beq C \approx - 0.09601 N^{3/2}, \quad N \to \infty \label{CN}\eeq
while for the physical value $N = 2$, 
\beq C \approx -0.04455, \quad N = 2 \eeq
The $N^{3/2}$ behaviour in Eq. (\ref{CN}) indicates a breakdown of the genus expansion of Ref. \onlinecite{SSLee}. Moreover, since this correction is parametrically larger than the tree level value, the existence of the large-$N$ limit of the theory is cast into doubt. In particular, it is not clear if there are higher loop graphs with even stronger divergences in the large-$N$ limit. Moreover, we expect contributions to the bosonic self-energy analytic in $q_y$ to be generated from kinematic regimes involving the whole Fermi-surface and not just the two Fermi patches. Such analytic contributions might also exhibit anomalous scaling with $N$.   

Note that there is no logarithmic dependence on $\Lambda/\mu$ in Eq. (\ref{deltaPires}), and so we have $z=3$ at this order. For the physical value of $N = 2$, the finite three-loop correction turns out to be rather small numerically.

\subsection{Fermion anomalous dimension}
\begin{figure}[h]
\begin{center}
\includegraphics*[width=3in]{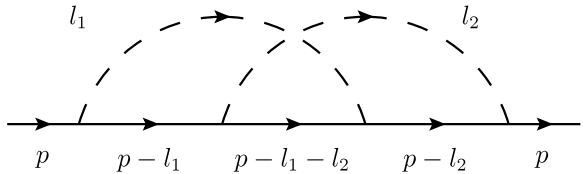}
\end{center}
\caption{Fermion self-energy at two loops.}
\label{Fig2loopfer}
\end{figure}

\begin{figure}
\begin{center}
\includegraphics*[width=4.5in]{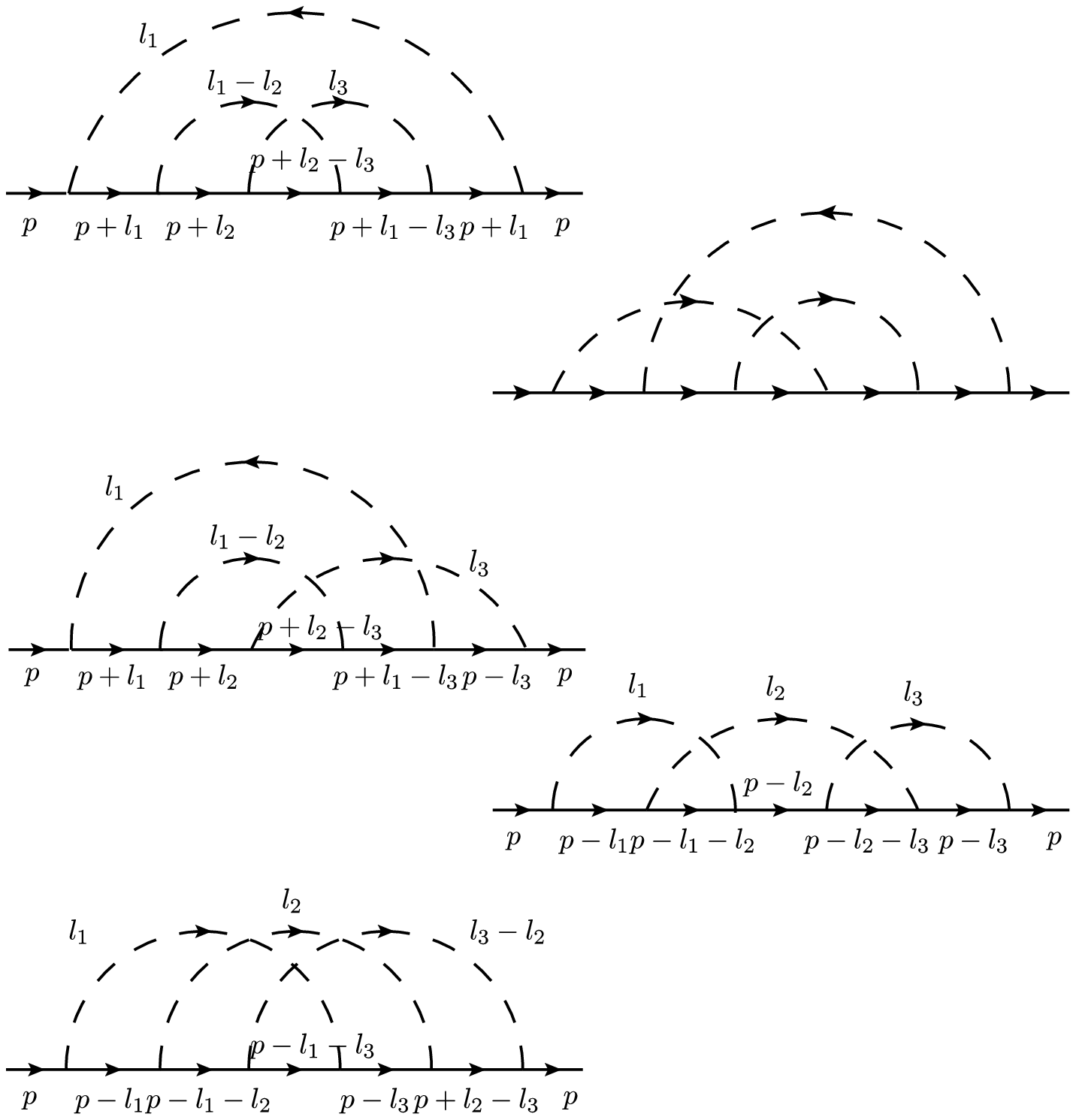}
\end{center}
\caption{Three loop fermion self-energy diagrams with no fermion loops}.
\label{Fig3loopfer}
\end{figure}

\begin{figure}[h]
\begin{center}
\includegraphics*[width=4in]{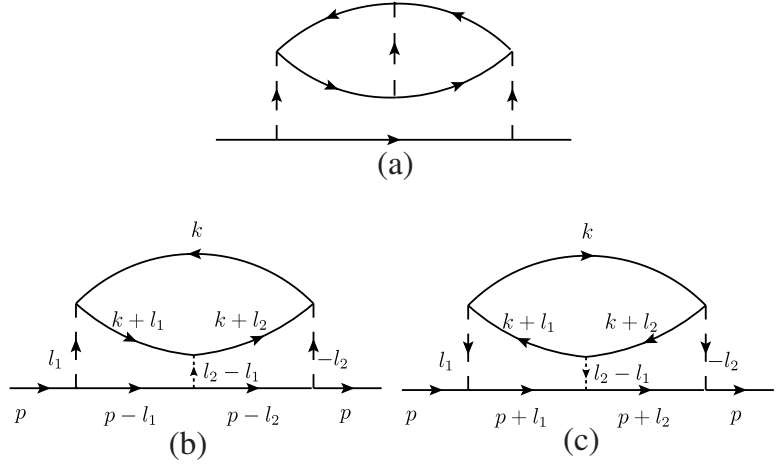}
\end{center}
\caption{Three loop fermion self-energy diagrams with one fermion loop}
\label{FigSE3loop}\end{figure}

The Feynman diagrams for the fermion self-energy up to three loop order are shown in Figs. \ref{Fig2loopfer}, \ref{Fig3loopfer} and \ref{FigSE3loop}. By reasons explained in the previous section, the diagrams in Figs. \ref{Fig2loopfer} and \ref{Fig3loopfer} vanish when the external frequency is zero and, hence, do not contribute to the fermion anomalous dimension. 

\begin{figure}
\begin{center}
\includegraphics*[width=2.5in]{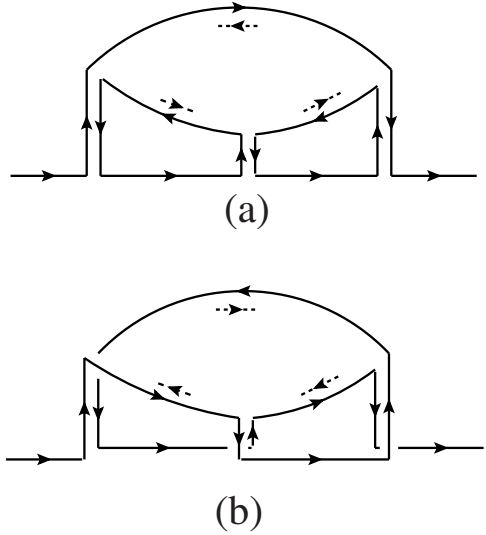}
\end{center}\caption{Double line representation of fermion self-energy diagrams in Figs. \ref{FigSE3loop}b,c, 
as in Fig.~\ref{FigAzLarkDouble}. The external fermions and the fermions inside the loop are assumed to come from opposite patches.} \label{DoubleSE}
\end{figure}

Thus, the only fermion self-energy diagrams that can give UV divergences are shown in Fig. \ref{FigSE3loop}. Actually, the diagram in Fig. \ref{FigSE3loop} a) is zero since the polarization correction in Fig. \ref{PiTwoLoop} vanishes. Thus, we only need to consider the two diagrams in Fig. \ref{FigSE3loop} b) and c). For these graphs to be UV divergent, the fermions running in the loop and the external fermions must come from different patches. The diagram in Fig. \ref{FigSE3loop} b) contains two loops in the double line representation (Fig. \ref{DoubleSE} a)) and is expected to be of order $1/N$, while the one in Fig. \ref{FigSE3loop} c) has no loops in the double line representation (Fig. \ref{DoubleSE} b)) and, hence, is expected to scale as $1/N^2$. 

A calculation presented in Appendix~\ref{app:comp} gives the $UV$ divergent contribution,
\bea \delta^{3b} \Sigma_+(\omega = 0, \vec{p}) &=& \lambda_+ \lambda_- J_b (p_x + p^2_y) \log\left(\frac{\Lambda_y}{ |p_x + p^2_y|^{1/2}}\right), \label{SE3loopb}\\
\delta^{3c} \Sigma_+(\omega = 0, \vec{p}) &=& \delta^{3c} \Sigma_+(\omega = 0, \vec{p}=0) + \lambda_+ \lambda_- J_c (p_x + p^2_y) \log\left(\frac{\Lambda_y}{|p_x + p^2_y|^{1/2}}\right)\label{SE3loopc}\eea
The constant $J_b$ is independent of $N$ and given numerically by,
\beq J_b \approx 0.1062 \label{Jb} \eeq
On the other hand, the constant $J_c$ is $N$-dependent. For $N= 2$ we obtain,
\beq J_c \approx -0.03795, \quad N = 2 \label{Jc2}\eeq
while in the large-$N$ limit,
\beq J_c \approx \frac{9}{4 \pi^2 N^2} \log^3 N, \quad N \to \infty \label{JcN} \eeq
Notice that there is no $1/N$ suppression in Eq. (\ref{SE3loopb}). A way to interpret this, is that the diagram is really of order $1/N$ (as the genus expansion predicts), however, it is a function of $N (p_x + p^2_y)$. Indeed, recall that the genus expansion assumes $N (p_x + p^2_y) \sim 1$. However, the $UV$ divergent piece of the diagram cannot depend on the magnitude of $p_x + p^2_y$ and is valid for any external momentum or frequency. On the other hand, the infrared scale under the log is expected to become $\omega^{1/3}$ once $\omega \gg N^{3/2} |p_x + p^2_y|^{3/2}$. Also observe that up to a logarithmic enhancement, the non-planar diagram \ref{FigSE3loop} c) (\ref{DoubleSE} b)) is of order $1/N^2$, as expected from the genus expansion. 

Note that the $UV$ divegence in Eqs. (\ref{SE3loopb}), (\ref{SE3loopc})  is logarithmic, as expected from power counting, and comes from a region where both internal momenta and frequencies diverge in accordance with the scaling (\ref{scaling}). This is unlike the anomalous linear divergences of the Aslamazov-Larkin diagrams that occur when the internal momenta $q_y$ are of order of external momenta, while internal $q_x$, $q_{\tau}$ diverge.

Thus, to three loop order,
\beq \delta^{3} \Sigma_+(\omega = 0, \vec{p}) = \delta^{3} \Sigma_+(\omega = 0, \vec{p}=0) + \lambda_+ \lambda_- J (p_x + p^2_y) \log\left(\frac{\Lambda_y}{|p_x + p^2_y|^{1/2}}\right)\label{SE3loop}\eeq
\bea J = J_b + J_c \approx \left\{\begin{array}{cc} 0.06824& \quad N = 2\\0.10619& \quad N = \infty\end{array}\right.\eea  
Although the self-energy correction (\ref{SE3loop}) is not parameterically suppressed compared to the bare value even when $N = \infty$, it appears to be  suppressed numerically. Thus, we may estimate,
\beq Z_\psi = 1 - \lambda_+ \lambda_- J \log \Lambda/\mu \nn \eeq
\beq \eta_\psi = \lambda_+ \lambda_- J = \pm 0.06824 \label{etafval} \eeq
where the upper sign refers to the Ising-nematic transition and the lower sign to the spin-liquid and we have used the value of $J$ at $N = 2$. 

\subsection{Fermi surface shift}
We now evaluate the coefficient $\alpha$, Eq. (\ref{alphadef}), associated with the renormalization of chemical potential $\delta$ away from criticality. This coefficient can be obtained from the insertion of the $\phi^2$ operator into the two-point fermion Green's function at criticality. By setting all external frequencies to zero, we find that at three loop order the only $UV$ divergent contribution can originate from the diagrams in Figs. \ref{FigSE3loop} b) c) with the $\phi^2$ operator inserted into the boson propagators. The details of the calculation are presented in appendix \ref{app:comp}. We find,
\beq \delta^3 \frac{\d \Sigma}{\d r} \stackrel{UV}{=} J_r e^2 \log \Lambda_y \eeq
with
\bea J_r = 0.00208, \quad N = 2 \nn\\
J_r \sim O\left(\frac{1}{N^3}\right), \quad N \to \infty \label{Jrnum}\eea
Absorbing this divergence into the chemical potential,
\beq Z_{r \delta} = J_r \log \Lambda_y/\mu \eeq
and 
\beq \alpha = J_r \eeq
Thus, the $\phi^2$ operator mixes with the $\psi^{\dagger} \psi$ operator. If the dynamical critical exponent $z = 3$, this  leads to a logarithmic divergence of the compressibility, Eq. (\ref{compr3}). Note that the magnitude of the mixing $\alpha$ is suppressed in the large $N$ limit and is also numerically small for $N = 2$.

\section{Conclusions}
\label{sec:conc}

This paper has presented the scaling properties of the field theory in Eq.~(\ref{L})
which describes a number of problems involving the breakdown of Landau Fermi liquid
theory at all points on a two dimensional Fermi surface. The main motivation was provided
by the quantum phase transition caused by the onset of Ising-nematic order,
which reduces the point-group symmetry from square to rectangular.
However our theory also directly applies or can be generalized to breaking of other point-group and/or time-reversal
symmetries, and these were described in Section~\ref{sec:model}. One of these cases is the
``circulating current'' order parameter of Simon and Varma \cite{vojtaprl,varma,berg}.
Apart from applications to quantum critical points, our theory also described non-Fermi liquid
phases associated with spin liquids \cite{hermele,senthil} or algebraic charge liquids \cite{rkk2,su2,qi}, which
have Fermi surfaces coupled to U(1) gauge fields.

Our critical theory was formulated in terms of a time-reversed pair of patches on the Fermi surface,
centered at the wavevectors $\pm \vec{k}_0$ (see Fig.~\ref{fig:fs}). The value of $\vec{k}_0$ was determined
by requiring that the tangent to the Fermi surface at $\vec{k}_0$ be parallel to the wavevector $\vec{q}$ carried by the
order parameter insertion in the correlation function being computed. However, in general, there is nothing special about
the point $\vec{k}_0$, and neighboring points on the Fermi surface should behave in a similar manner.
This key feature was implemented in our theory by the rotational symmetry discussed in Section~\ref{sec:rot}, 
and the identities (\ref{slideb},\ref{slidef}), which show that the Green's function remains invariant as we
move along the Fermi surface.

We emphasize that although we have critical theories associated with every pair of points
on the Fermi surface, the Lagrangian (\ref{L}) and all the fields are 2+1 dimensional {\em i.e.\/}
$\phi$ and $\psi_{\sigma}$ are integrated over arbitrary functions of $x$, $y$, and $\tau$. Thus, as we noted
earlier, our approach
and results 
differ from studies using a `tomographic' representations of the Fermi surface, in which every point on the 
Fermi surface is described by a 1+1 dimensional field theory.\cite{Luther,Marston,KwonMarston,haldane,NetoFradkin,LF07,LFBFO06}
 Our 2+1 dimensional representation
leads to a redundancy in our description of the degrees of freedom, and the identities 
of Section~\ref{sec:rot} ensure the consistency of this redundant description.

Our main results include the 
scaling relations for the order parameter susceptibility in Eq.~(\ref{Dscal}), and for the fermion Green's function
in Eq.~(\ref{Gscal}). These are associated with only two independent exponents, the dynamic scaling exponent $z$,
and the fermion anomalous dimension $\eta_\psi$. The correlation length exponent $\nu$
was given by exact scaling relation in Eq.~(\ref{nuz}), while the susceptibility exponent $\gamma = 1$.
For the spin-liquid case, Fermi liquid arguments were made  \cite{HALPERIN,KIM94,ybkim2,stern1,stern2,read}  
suggesting that $z=3$; we found $z=3$ to three loop order in Section~\ref{sec:aetl}, although we did not prove
this to all orders, and our scaling theory is compatible with a general value of $z$. 
Our three loop 
computation also gave a non-zero value of $\eta_\psi$, with opposite signs for the Ising-nematic
and spin-liquid cases. In the case of the nematic transition, a non-zero positive $\eta_\psi$ implies the suppression
of the electron tunneling  density of states, Eq. (\ref{DOS2}). Another striking effect that we find for the case
of a nematic transition is the power law divergence of the compressibility for $z > 3$, which turns into a logarithmic divergence
if $z = 3$. 

Our scaling results were expressed in terms of correlators of the fermionic field $\psi_{+ \sigma}$ carrying
momentum $\vec{q}$ as measured from the point $\vec{k}_0$ from the Fermi surface, implying from (\ref{psic})
that the electron $c_{\sigma}$ has momentum $\vec{k}_0 + \vec{q}$ (and similarly for $\psi_{-\sigma}$).
However, note that (after appropriate rescaling of momenta, and for a circular Fermi surface)
$|\vec{k} | - k_F \approx q_x + q_y^2$.
Thus the identity (\ref{slidef}) implies that the scaling function (\ref{Gscal}) for the two-point fermion
Green's function depends only on $|\vec{k} | - k_F$. 
This is similar to the dependence found in other
treatments {\em e.g.\/} in the recent critical theories \cite{sslee3,schalm,lmv,ssdenef} 
obtained by applying the AdS/CFT duality 
to fermions propagating near a Reissner-Nordstrom black hole. The latter theories, in their current
classical gravity formulation, find \cite{lmv} $\eta_\psi = 0$.

It is also interesting to compare the structure of the critical theory in the AdS/CFT framework
to that found here. We have an infinite set of 2+1 dimensional field theories labeled by pairs of momenta
on a one-dimensional Fermi surface {\em i.e.\/} a $\mathbb{S}^1 /\mathbb{Z}_2$ set of 2+1 dimensional field theories.
In the low-energy limit, the AdS/CFT approach yields \cite{lmv} a AdS$_2 \times \mathbb{R}^2$ geometry: this can be interpreted
as an infinite set of chiral 1+1 dimensional theories labeled by a $\mathbb{R}^2$ set of two-dimensional momenta
$\vec{k}$. It is notable, and perhaps significant, that both approaches have an emergent dimension not found in the underlying degrees of freedom.
We began with a 2+1 dimensional Hamiltonian, and ended up with a $\mathbb{S}^1 /\mathbb{Z}_2$ set of 
2+1 dimensional field theories. In AdS/CFT, there is the emergent radial direction representing energy scale.
These emergent dimensions imply redundant descriptions, and require associated consistency conditions:
we explored such consistency  conditions in Section~\ref{sec:rot}, while in AdS/CFT the
consistency 
conditions are Einstein's equations representing the renormalization group flow under changes of energy scale.
It would be interesting to see if fluctuations about the classical gravity theory
yield corrections to the AdS$_2 \times \mathbb{R}^2$ geometry which clarify the connection to our theory.

In the analysis of the spin-liquid problem, Ref.~\onlinecite{SSLee} considered a single patch of the Fermi surface,
and argued that the $1/N$ expansion should be organized by the genus of the Feynman graph
(after the propagators are written in a suitable double line representation, and the graph is interpreted as
lying on a two-dimensional surface). In our two-patch theory here, we have shown that this genus counting is violated.
This is the implication of the $N^{3/2}$ dependence of the boson self-energy in Eq.~(\ref{deltaPires}). In fact, at present, it is not clear how to take the large-$N$ limit of the theory. On the other hand, for the physical value $N = 2$, we found that the higher loop contributions are numerically small, which suggests that the critical exponents are close to the 
Hertz mean-field values. However, because the loop-wise expansion does not possess even a formal expansion parameter, it is not clear if there is a systematic way to extract corrections to the mean-field exponents. Thus, our value of the fermion anomalous dimension $\eta_\psi$, Eq. (\ref{etafval}), should be regarded as an estimate only.

\acknowledgements

We thank E.~Altman, G.~Baskaran, A.~Chubukov, E.~Fradkin, B.~I.~Halperin, S.~Hartnoll, Y.-B.~Kim, S.~A.~Kivelson, 
S.-S.~Lee, W.~Metzner, Y.~Oreg, D.~Scalapino, A.~Schiller, T.~Senthil, 
R. Shankar (IMSc), R.~Shankar (Yale), A.~Stern, L.~Taillefer, and C.~Xu for useful discussions.
This research was supported by the National Science Foundation under grant DMR-0757145, by the FQXi
foundation, and by a MURI grant from AFOSR.

\appendix

\section{Decoupling of non-collinear momenta}
\label{app:decoupling}
In this section we will argue that the fluctuations of the order parameter at non-collinear momenta effectively decouple. We focus for simplicity on the case of an Ising-nematic transition. We follow the 
standard Hertz approach, integrating out the fermions to obtain an effective action for $\phi$, 
\beq S[\phi] = \sum_{n=2}^{\infty} \frac{1}{n!} \int d^Dx_1 \ldots d^D x_n \Gamma^n(x_1, x_2, \ldots x_n) \phi(x_1)\phi(x_2)\ldots \phi(x_n)\label{SGamma}\eeq
The $n$-point effective vertex $\Gamma^n$ is given by,
\beq \Gamma^n(q_1,q_2,\ldots q_n) = \frac{N}{n} f^n(q_1,q_2,\ldots q_n) + \mathrm{permutations\,of\,} q_1,q_2,\ldots q_n \label{Gamman}\eeq
with 
\beq f^n(q_1,q_2,\ldots q_n) = \int \frac{dk_\tau d^2 k}{(2\pi)^3} \prod_{i = 0}^{n-1} \left[G(k + l_i) d_{\vec{k} + \frac{\vec{l}_i + \vec{l}_{i+1}}{2}}\right]\eeq
where $l_i = \sum_{j=1}^{i} q_j$. For now we work with ``undressed" propagators,
\beq G(\omega, \vec{k}) = \frac{1}{-i \omega + v_F(\theta) k}\eeq
with $k$ - the distance to the Fermi surface and $v_F(\theta)$ - the local Fermi velocity. As is well-known, for $\omega \ll v_F |\vec{q}|$ and $|\vec{q}| \ll k_F$ the two-point vertex has a Landau-damped form,
\beq \Gamma^2(\omega, \vec{q}) = N \left[\gamma(\hat{q}) \frac{|\omega|}{|\vec{q}|} + \frac{\vec{q}^2}{e^2} +r\right] \label{LandauDamp}\eeq
where the coefficient of the non-analytic term $\gamma(\hat{q}) =  {K d^2}/{(2 \pi v^2_F)}$ with the Fermi-surface curvature radius $K$, Fermi-velocity $v_F$ and form-factor $d$ evaluated at the point on the Fermi surface to which $\vec{q}$ is tangent. On the other hand, the coefficients of the analytic terms $r$ and $1/e^2$ come from the entire Fermi-surface.

If we truncate the series (\ref{SGamma}) at the quadratic order,
\beq S_2 = \frac{N}{2} \int \frac{d\omega d^2 \vec{q}}{(2 \pi)^3} \left[\gamma(\hat{q}) \frac{|\omega|}{|\vec{q}|} + \frac{\vec{q}^2}{e^2} +r\right] |\phi(\vec{q},\omega)|^2 \label{S2}\eeq
then at the critical point $r = 0$ the action (\ref{S2}) is invariant under the scale transformation,
\beq \phi(\vec{x}, \tau) \to s^{3/2} \phi(s \vec{x}, s^3 \tau) \label{scalHertz}\eeq
Note that here, in contrast to Eq.~(\ref{scaling}), all components of $\vec{q}$ are scaled in the same way as we are not studying the effects of fluctuations with collinear wave-vectors. We can regard the terms in Eq.~(\ref{SGamma}) with $n > 2$ as perturbations to the Hertz action (\ref{S2}). Hertz noted that if the effective vertices $\Gamma^n$ possess a regular expansion in frequencies and momenta, such that the corresponding operators can be represented as polynomials in the order parameter $\phi$ and its derivatives, then the perturbations with $n > 2$ are irrelevant due to the large effective dimensionality, $D_{eff} = d + z = 5$, with $d=2$ - spatial dimension and $z=3$ - the dynamical critical exponent. Indeed, the perturbation $\int d^2\vec{x} d\tau \phi^n(x)$ scales as $s^{3n/2-5}$ under (\ref{scalHertz}) (in the special case $n = 3$, the operator $\phi^3$ is actually prohibited by the $90^\circ$ lattice rotation symmetry. The lowest dimension local operators with three powers of $\phi$ that are allowed by symmetry are $\phi ((\d_x \phi)^2 - (\d_y \phi)^2)$ in the $d_{x^2-y^2}$ case and $\phi \d_x \phi \d_y \phi$ in the $d_{xy}$ case, which scale as $s^{3/2}$). 

However, due to the presence of low-energy excitations on the Fermi-surface there is no reason to expect that the effective vertices $\Gamma^n$ would possess a regular expansion in momenta. Indeed, we have already seen that the two-point vertex has the non-analytic Landau-damped form (\ref{LandauDamp}). As we now show, similar non-analyticities occur in the higher order vertices.

Let us estimate the vertices (\ref{Gamman}) when the external frequencies and momenta obey the Hertz scaling (\ref{scalHertz}), $\omega \sim |\vec{q}|^3$, $\vec{q} \to 0$. In this regime,
\beq f^n(q_1,q_2,\ldots q_n) = \int \frac{d k_\tau dk d \theta}{(2 \pi)^3} \left|\frac{d \vec{k}_F}{d \theta}\right| d(\theta)^n \prod_{i =0}^{n-1} \frac{1}{-i (k_\tau + l_{i \tau}) + v_F(\theta)(k+ \hat{v}_F(\theta) \cdot \vec{l}_i)}\eeq
Let us perform the integral over $k$. Observe that if $|k_\tau| > \Omega$, with $\Omega = \max_i |l_{i \tau}|$ then the integral vanishes as all the poles of the integrand are in the same half-plane. Thus, the range of the internal frequency is limited by the external ones. With this in mind,
\bea f^n(q_1,q_2,\ldots q_n) &=& i \int_{|k_\tau| < \Omega} \frac{d k_\tau}{2 \pi} \int \frac{d \theta}{2 \pi} \left|\frac{d \vec{k}_F}{d \theta}\right| \frac{d(\theta)^n}{v_F(\theta)} \nonumber \\
&~&~~~~~~~\times \sum_{j=0}^{n-1} \vartheta(k_\tau + l_{j\tau}) \prod_{i =0, i \neq j}^{n-1} \frac{1}{-i (l_{i\tau} - l_{j \tau}) + \hat{v}_F(\theta) \cdot (\vec{l}_i - \vec{l}_j)}, \label{Gammanlilj}\eea
where we have used the symbol $\vartheta$ for the step function, to avoid confusion with the angular variable
$\theta$.
Now, since $q_\tau \sim |\vec{q}|^3/(\gamma e^2) \ll v_F |\vec{q}|$, for general $\theta$ we can ignore the frequency dependence in the denominator of Eq.~(\ref{Gammanlilj}). Then the angular integration yields a factor of ${\cal O}(1)$ and the integral over $k_\tau$ yields a factor of external frequency, so that
\beq \Gamma^n(q_1,q_2,\ldots q_n) \sim \frac{q_\tau}{|\vec{q}|^{n-1}}\label{Gammaest}\eeq
Note that the momentum dependence in Eq.~(\ref{Gammaest}) is far from analytic. Also, note that for $n=2$ the result is consistent with the standard Landau damping.

The only possible caveat to the estimate (\ref{Gammaest}) is associated with regions of angular integration where $\hat{v}_F(\theta) \cdot (\vec{l}_i - \vec{l}_j) \to 0$, i.e some combination of external momenta becomes tangent to the Fermi surface. Then the angular integration acquires poles just off the real axis, with the imaginary parts of the poles provided by the frequency dependence in the denominator of Eq.~(\ref{Gammanlilj}). As long as the real parts of the poles do not coalesce, i.e no two momenta $\vec{l}_i - \vec{l}_j$ and $\vec{l}_{i'} - \vec{l}_j$ are collinear, the angular integration still yields a factor of ${\cal O}(1)$ and the estimate (\ref{Gammaest}) remains correct. This is the regime that we are considering in the present appendix. The rest of the paper is devoted to the opposite limit, where all the external momenta are nearly collinear and the angular integral in Eq.~(\ref{Gammanlilj}) is dominated by the vicinity of two antipodal points on the Fermi surface to which the external momenta are tangent. This observation motivates the introduction of the two patch theory in Section \ref{sec:model} and all the subsequent development of the present work.

Returning to the non-collinear regime, upon combining Eq.~(\ref{Gammaest}) with the Hertz scaling (\ref{scalHertz}), we conclude that the $n$-th term in the series (\ref{SGamma}) scales as $s^{n/2 -1}$. Therefore, all terms with $n > 2$ represent non-local {\it irrelevant} perturbations, which confirms that the fluctuations with non-collinear momenta decouple.

We would like to point out that the argument above still holds if one dresses the fermion propagator by the one-loop self-energy,  $\Sigma(\omega, k) \sim -i \mathrm{sgn} \omega |\omega|^{2/3}$. This modifies the frequency dependence in the denominator of Eq.~(\ref{Gammanlilj}) via,  $-i (l_{i\tau} - l_{j \tau}) \to \Sigma(k_\tau + l_{i \tau}) - \Sigma(k_\tau + l_{j \tau})$.  However, since $\Sigma(\omega) \ll v_F |\vec{q}|$ for  typical $\omega \sim |\vec{q}|^3$, the estimate (\ref{Gammaest}) is still correct.

\section{Computations of Feynman diagrams}
\label{app:comp}

Here we provide some details of the computations of the diagrams in Section~\ref{sec:aetl}.

\subsection{Boson self-energy}
We begin by evaluating the two-loop polarization correction in Fig. \ref{PiTwoLoop},
\beq \delta^2 \Pi(q) = N \sum_s \int \frac{dp_{\tau} d^2p}{(2 \pi)^3} \frac{dl_{\tau} d^2 l}{(2 \pi)^3} D(l) G_s(p) G_s(p+q) G_s(p-l) G_s(p+q-l)\eeq
The contributions to the integral from the two patches are equal. Thus, integrating over $p_x$, $l_x$ we obtain,
\bea \delta^2 \Pi(q) &=& 2 N \int \frac{dp_{\tau} dp_y}{(2 \pi)^2} \frac{dl_\tau dl_y}{(2\pi)^2} D(l) \frac{\theta(p_\tau)-\theta(p_\tau + q_\tau)}{\frac{i c_f}{N} (\{p\} - \{p + q\}) +  2 q_y p_y + q_x+ q^2_y}\nn\\&\times& \frac{\theta(l_\tau-p_\tau) - \theta(l_\tau-p_\tau-q_\tau)}{\frac{i c_f}{N} (\{l - p - q\} - \{l-p\}) + 2 q_y (p_y - l_y) + q_x + q_y^2}\eea
where here and below we use the notation $\{p\} = \mbox{sgn}(p_\tau) |p_\tau|^{2/3}$. We observe that the poles of the $p_y$ integral are always in the same half-plane. Thus, $\delta^2 \Pi(q) = 0$. This is consistent with Ref. \onlinecite{KIM94}, which found that the two loop corrections to Eq. \ref{Pi0} are suppressed by factors of $|\omega|^{2/3}$ or $|\omega|/|q_y| \sim |\omega|^{2/3}$.  

Now, let us proceed to compute the Aslamazov-Larkin diagrams, Fig. \ref{FigAzLark}. We begin by evaluating the three point-function $f_s(q, l, -(l+q))$ in Eq.~(\ref{f}). 
Note that $f_-(q,l,-(l+q)) = f_+(P_x q, P_x l , -P_x(l+q))$ where $P_x (k_0, k_x, k_y) = (k_0, -k_x, k_y)$. The calculation of $f$ is simplified when $q_{\tau} = 0$. Then, performing the integral over $p_x$ and, subsequently, $p_y$, in Eq. (\ref{f}),
\bea f_+(q, l, -(l+q)) &\stackrel{q_{\tau} = 0}{=}& \int \frac{d p_\tau dp_y}{(2 \pi)^2} \frac{i (\theta(p_\tau + l_\tau) -\theta(p_\tau))}{\frac{i c_f}{N} (\{p+l\} - \{p\}) - l_x - 2 l_y p_y  - l_y^2} \times 
\nn\\&&\frac{1}{\frac{i c_f}{N} (\{p+l\} - \{p\}) - q_x -l_x - 2 (q_y + l_y) p_y  + q^2_y - l^2_y}\nn\\
&=& \frac{1}{2 q_y} \int \frac{d p_\tau}{2 \pi} \frac{|\theta(p_\tau + l_\tau) - \theta(p_\tau)| (\theta(l_y) - \theta(q_y + l_y))}{ \frac{-i c_f}{N} (\{p+l\} - \{p\}) + l_x - \frac{q_x}{q_y} l_y + l_y (q_y + l_y)}\eea
Thus,
\bea \delta^3 \Pi(q_{\tau} = 0, \vec{q}) &=& \frac{\lambda_+ \lambda_- N^2}{4 q^2_y} \int \frac{dl_\tau d^2\vec{l}}{(2 \pi)^3} \frac{dp_\tau}{2\pi} \frac{dp'_\tau}{2 \pi} D(l) D(l+q) \times\nn\\
 && \frac{|\theta(p_\tau + l_\tau)-\theta(p_\tau)| |\theta(p'_\tau + l_\tau) - \theta(p'_\tau)| |\theta(l_y) - \theta(l_y + q_y)|}{\frac{-i c_f}{N} (\{p+l\} - \{p\}) + l_x - \frac{q_x}{q_y} l_y + l_y (q_y + l_y)}\times\nn\\
&&  \bigg(\frac{1}{\frac{-i c_f}{N} (\{p'+l\} - \{p'\}) - l_x +\frac{q_x}{q_y} l_y - l_y (q_y + l_y)}\nn \\
&-& \frac{1}{\frac{-i c_f}{N} (\{p'+l\} - \{p'\}) - l_x +\frac{q_x}{q_y} l_y + l_y (q_y + l_y)}\bigg) + (q \to - q)\eea 
Finally, integrating over $l_x$,
\bea \delta^3 \Pi(q_{\tau} = 0, \vec{q}) &=& \frac{\lambda_+ \lambda_- N^2}{4 q^2_y} \int \frac{dl_\tau dl_y}{(2 \pi)^2} \frac{dp_\tau}{2\pi} \frac{dp'_\tau}{2 \pi} D(l) D(l+q) \times\nn\\
 && i \mbox{sgn}(l_\tau) |\theta(p_\tau + l_\tau)-\theta(p_\tau)| |\theta(p'_\tau + l_\tau) - \theta(p'_\tau)| |\theta(l_y) - \theta(l_y + q_y)| \times\nn\\
 && \bigg(\frac{1}{\frac{-i c_f}{N} (\{p+l\} - \{p\} + \{p'+l\} - \{p'\})}\nn\\ &-& \frac{1}{\frac{-i c_f}{N} (\{p+l\} - \{p\} + \{p'+l\} - \{p'\}) + 2 l_y (q_y + l_y)}\bigg) + (q \to -q)\nn\\\eea
The integral is invariant under $q \to -q$. Moreover, the integrals in the regions $l_0 > 0$ and $l_0 < 0$ are related by complex conjugation. Thus,
\bea \delta^3 \Pi(q_{\tau} = 0, \vec{q}) &=& -\frac{\lambda_+ \lambda_- N}{q^2_y} \int_0^{\infty} \frac{d l_\tau}{2 \pi} \int_0^{l_\tau} \frac{dp_\tau}{2 \pi} \int_0^{l_\tau} \frac{d p'_\tau}{2 \pi} \int_0^{|q_y|} \frac{d l_y}{2 \pi} \frac{1}{c_b \frac{l_\tau}{l_y} + \frac{l^2_y}{e^2}}\frac{1}{c_b \frac{l_\tau}{|q_y|-l_y} + \frac{(|q_y|-l_y)^2}{e^2}} \times\nn\\
 &&  \bigg(\frac{1}{c_f ((l-p)^{2/3}_\tau +p^{2/3}_\tau + (l-p')^{2/3}_\tau + p'^{2/3}_\tau)}\nn\\ &-& \frac{c_f ((l-p)^{2/3}_\tau +p^{2/3}_\tau + (l-p')^{2/3}_\tau + p'^{2/3}_\tau) }{c^2_f((l-p)^{2/3}_\tau + p^{2/3}_\tau + (l-p')^{2/3}_\tau + p'^{2/3}_\tau)^2 + 4 N^2 l^2_y (|q_y| - l_y)^2}\bigg) \eea 
Notice that the integral over $l_y$ is bounded by the external momentum $q_y$. This leads to a violation of the naive power counting, which would predict that each diagram in Fig. \ref{FigAzLark} has a superficial degree fo divergence $\Lambda^2_y \sim \Lambda^{2/3}_\tau$. Instead, we find that for $l_\tau \to \infty$, the two diagrams behave as,
\beq \delta^{3a} \Pi(0, \vec{q}) = -\delta^{3b} \Pi(0, \vec{q}) \sim - \lambda_+ \lambda_- N |q_y| \left(\frac{\Lambda_\tau}{e^4}\right)^{1/3}\eeq
(In reality, the divergence is cut once we exit the two patch regime where the momentum $l_x \ll l_y$. This occurs when $l_x \sim l^{2/3}_\tau$ becomes of order $l_y$. However, for the Aslamazov-Larkin diagrams the internal momentum $l_y$ is controlled by external momentum $q_y$. Hence, $\Lambda_\tau \sim q^{3/2}_y$ and $\delta^{3a} \Pi = - \delta^{3b} \Pi \sim q^{3/2}_y$, as found in Ref. \onlinecite{RPC06}).

However, as expected for problems involving a boson field coupled to the charge sector of the Fermi-surface, the divergence cancels when we add the two diagrams. In fact, for $N \gg 1$, the divergence is cut-off at $\frac{c_f}{N} l^{2/3}_\tau \sim q^2_y$, {\em i.e.\/} 
\beq l_\tau \sim N^{3/2} q^3_y/e^2 \label{rangesat} \eeq
 so that
\beq \delta^3 \Pi(0, \vec{q}) \sim -\lambda_+ \lambda_- N^{3/2} \frac{q^2_y}{e^2} \label{deltaPiest}\eeq
Note that the result is parameterically larger in the large-$N$ limit than the bare boson polarization, Eq. (\ref{L}) (although it has the same scaling as the bare term). Also observe that the sign of the contribution (\ref{deltaPiest}) is positive for the spin-liquid 
and negative for the Ising-nematic transition. 

One may ask whether the enhancement in (\ref{deltaPiest}) is an artifact of taking $q_{\tau} = 0$. However, since the integral in Eq. (\ref{deltaPiest}) is saturated in the region (\ref{rangesat}), we expect the result (\ref{deltaPiest}) to be valid for, $q_{\tau} \ll N^{3/2} q^3_y/e^2$, which is certainly satisfied by the typical bosonic momenta $q_{\tau} \sim q^3_y/e^2$. 

We can compute the proportionality factor in Eq. (\ref{deltaPiest}) in the large-$N$ limit. Changing variables to $l_\tau = \left(\frac{N}{c_f}\right)^{3/2} |q_y|^3 \bar{l}_\tau$, $p_\tau = l_\tau x$, $p'_\tau = l_\tau x'$, $l_y = |q_y| y$,
\beq \delta^3 \Pi(0, \vec{q}) = C \lambda_+ \lambda_-  \frac{q^2_y}{e^2} \label{Piprop}\eeq
\bea C &=&  - \frac{2^{5/2} 3^{3/4} N^{3/2}}{\pi} \int_0^{\infty} d\bar{l}_\tau \int_0^1 dx \int_0^1 dx' \int_0^1 dy \frac{\bar{l}^{4/3}_\tau y^3 (1-y)^3}{(\bar{l}_\tau + \left(\frac{2}{N \sqrt{3}}\right)^{3/2} y^3) (\bar{l}_\tau + \left(\frac{2}{N \sqrt{3}}\right)^{3/2} (1-y)^3)}\nn\\ &\times&
 \frac{1}{A (A^2 \bar{l}^{4/3}_\tau + 4 y^2 (1-y)^2)}\eea  
with,
\beq A = x^{2/3} + (1-x)^{2/3} + x'^{2/3} + (1-x')^{2/3}\eeq
For $N \gg 1$, the integral over $\bar{l}_\tau$ is saturated in the region $\bar{l}_\tau \sim 1$, so,
\beq C \approx - \frac{2^{5/2} 3^{3/4} N^{3/2}}{\pi} \int_0^{\infty} \frac{d\bar{l}_\tau}{\bar{l}^{2/3}_\tau} \int_0^1 dx \int_0^1 dx' \int_0^1 dy \frac{y^3(1-y)^3}{A (A^2 \bar{l}^{4/3}_\tau + 4 y^2 (1-y)^2)}\eeq
After a change of variables, $z = A \bar{l}^{2/3}_\tau/(2 y(1-y))$,
\beq C \approx - \frac{3^{7/4} N^{3/2}}{\pi} \int_0^{\infty} \frac{dz}{z^{1/2} (z^2 + 1)} \int_0^1 dy \, y^{3/2} (1-y)^{3/2} \int_0^1 dx \int_0^1 dx' \frac{1}{A^{3/2}} = \frac{3^{11/4} \pi N^{3/2}}{2^{15/2}} \int_0^1 dx \int_0^1 dx' \frac{1}{A^{3/2}} \eeq
The integral over $x,x'$ can be evaluated numerically,
\beq \int_0^1 dx \int_0^1 dx' \frac{1}{A^{3/2}} = 0.269653\eeq
so that
\beq C \approx - 0.09601 N^{3/2}, \quad N \to \infty\eeq
We may also compute the constant $C$ in Eq. (\ref{Piprop}) for the physical value $N = 2$,
\beq C \approx -0.04455 \eeq

\subsection{Fermion self-energy}
We next compute the three loop corrections to the fermion self-energy in diagrams Fig. \ref{FigSE3loop} b), c):
\bea \delta^{3b} \Sigma(p_\tau=0,\vec{p}) &=& N \lambda^3_+ \lambda^3_- \int \frac{dk_\tau d^2 k}{(2 \pi)^3} \frac{dl_{1 \tau} d^2 l_1}{(2 \pi)^3} \frac{dl_{2 \tau} d^2 l_2}{(2 \pi)^3} G_+(p-l_1) G_+(p-l_2) G_-(k) G_-(k+l_1) G_-(k+l_2)\nn\\&\times&D(l_1)D(l_2)D(l_1-l_2)\eea  
\bea \delta^{3c} \Sigma(p_\tau=0,\vec{p}) &=& N \lambda^3_+ \lambda^3_- \int \frac{dk_\tau d^2 k}{(2 \pi)^3} \frac{dl_{1 \tau} d^2 l_1}{(2 \pi)^3} \frac{dl_{2 \tau} d^2 l_2}{(2 \pi)^3} G_+(p+l_1) G_+(p+l_2) G_-(k) G_-(k+l_1) G_-(k+l_2)\nn\\&\times&D(l_1)D(l_2)D(l_1-l_2)\eea
Integrating over $l_{1x}$ and $l_{2x}$ we obtain,
\bea \delta^{3b} \Sigma(p_\tau=0,\vec{p}) = - N \lambda_+ \lambda_- \int \frac{dk_\tau d^2 k}{(2 \pi)^3} \frac{dl_{1 \tau} d l_{1y}}{(2 \pi)^2} \frac{dl_{2 \tau} dl_{2y}}{(2 \pi)^2}
 D(l_1) D(l_2) D(l_1 - l_2) \frac{1}{-\frac{i c_f}{N} k^{2/3}_\tau + \delta^-_k} \nn\\ \frac{\theta(l_{1 \tau} + k_\tau) - \theta(-l_{1\tau})}{-\frac{i c_f}{N} ((l_1 + k)_{\tau}^{2/3}+ l^{2/3}_{1 \tau}) + \delta^-_k + 2 (k+p)_y l_{1y} - \delta^+_p} 
 \frac{\theta(l_{2 \tau} + k_\tau) - \theta(-l_{2\tau})}{-\frac{i c_f}{N} ((l_2 + k)_{\tau}^{2/3}+ l^{2/3}_{2 \tau}) + \delta^-_k + 2 (k+p)_y l_{2y} - \delta^+_p} \nn\\ 
\eea
\bea \delta^{3c} \Sigma(p_\tau=0,\vec{p}) &=& - N \lambda_+ \lambda_- \int \frac{dk_\tau d^2 k}{(2 \pi)^3} \frac{dl_{1 \tau} d l_{1y}}{(2 \pi)^2} \frac{dl_{2 \tau} dl_{2y}}{(2 \pi)^2} D(l_1) D(l_2) D(l_1 - l_2)\frac{1}{-\frac{i c_f}{N} k^{2/3}_\tau + \delta^-_k}\nn\\
&&\frac{\theta(l_{1 \tau} + k_\tau) - \theta(-l_{1\tau})}{-\frac{i c_f}{N} ((l_1 + k)_{\tau}^{2/3}+ l^{2/3}_{1 \tau}) + \delta^-_k + 2 (k+p)_y l_{1y} + 2 l^2_{1y}+ \delta^+_p}\nn\\
&&\frac{\theta(l_{2 \tau} + k_\tau) - \theta(-l_{2\tau})}{-\frac{i c_f}{N} ((l_2 + k)_{\tau}^{2/3}+ l^{2/3}_{2 \tau}) + \delta^-_k + 2 (k+p)_y l_{2y} + 2 l^2_{2y}+ \delta^+_p}\nn\\\eea
where $\delta^\pm_p = \pm p_x + p^2_y$. Note the cancellation of the fermi-surface curvature terms $l^2_{1y,2y}$ in the ``planar graph" $\delta^{3b} \Sigma$.

We can reduce the integration range to $k_\tau > 0$, as the region $k_\tau < 0$ is related by complex conjugation. There are then four different kinematic regimes: i) $l_{1 \tau} > 0,\, l_{2 \tau} > 0$,\, ii) $l_{1 \tau} < -k_{\tau},\, l_{2 \tau} > 0$,\, iii) $l_{1 \tau} > 0,\, l_{2 \tau} < - k_\tau$,\, iv) $l_{1 \tau} < -k_\tau, \, l_{2 \tau} < - k_\tau$. The integral over $k_x$ in the regime i) vanishes as all the poles are in the same half-plane. The regimes ii) and iii) are related by $l_1 \leftrightarrow l_2$. Thus, 
\bea &&\delta^{3b} \Sigma(p_\tau=0,\vec{p}) = - N \lambda_+ \lambda_- \bigg[- 2 \int_0^{\infty} \frac{dk_\tau}{2\pi} \int \frac{d^2 k}{(2 \pi)^2} \int_{k_\tau}^{\infty} \frac{dl_{1 \tau}}{2 \pi} \int_0^{\infty} \frac{dl_{2\tau}}{2 \pi} \int \frac{d l_{1y} dl_{2y}}{(2 \pi)^2}\nn\\
 &&  D(l_1) D(l_2) D(l_{1\tau} + l_{2 \tau}, l_{1y} - l_{2y})  \frac{1}{-\frac{i c_f}{N} k^{2/3}_\tau + \delta^-_k} \nn\\
&&\frac{1}{\frac{i c_f}{N} ((l_1 - k)_{\tau}^{2/3}+ l^{2/3}_{1 \tau}) + \delta^-_k + 2 (k+p)_y l_{1y} - \delta^+_p} 
 \frac{1}{-\frac{i c_f}{N} ((l_2 + k)_{\tau}^{2/3}+ l^{2/3}_{2 \tau}) + \delta^-_k + 2 (k+p)_y l_{2y} - \delta^+_p}
\nn\\
&+& \int_0^{\infty} \frac{dk_\tau}{2\pi} \int \frac{d^2 k}{(2 \pi)^2} \int_{k_\tau}^{\infty} \frac{dl_{1 \tau}}{2 \pi} \int_{k_\tau}^{\infty} \frac{dl_{2\tau}}{2 \pi} \int \frac{d l_{1y} dl_{2y}}{(2 \pi)^2} D(l_1) D(l_2) D(l_1-l_2)  \frac{1}{-\frac{i c_f}{N} k^{2/3}_\tau + \delta^-_k}\nn\\
&&\frac{1}{\frac{i c_f}{N} ((l_1 - k)_{\tau}^{2/3}+ l^{2/3}_{1 \tau}) + \delta^-_k + 2 (k+p)_y l_{1y} - \delta^+_p} 
\frac{1}{\frac{i c_f}{N} ((l_2 - k)_{\tau}^{2/3}+ l^{2/3}_{2 \tau}) + \delta^-_k + 2 (k+p)_y l_{2y} - \delta^+_p}
\bigg]\nn\\
&& + h.c.
\eea 
\bea &&\delta^{3c} \Sigma(p_\tau=0,\vec{p}) = - N \lambda_+ \lambda_- \bigg[- 2 \int_0^{\infty} \frac{dk_\tau}{2\pi} \int \frac{d^2 k}{(2 \pi)^2} \int_{k_\tau}^{\infty} \frac{dl_{1 \tau}}{2 \pi} \int_0^{\infty} \frac{dl_{2\tau}}{2 \pi} \int \frac{d l_{1y} dl_{2y}}{(2 \pi)^2}\nn\\
 &&  D(l_1) D(l_2) D(l_{1\tau} + l_{2 \tau}, l_{1y} - l_{2y})  \frac{1}{-\frac{i c_f}{N} k^{2/3}_\tau + \delta^-_k} \nn\\
&&\frac{1}{\frac{i c_f}{N} ((l_1 - k)_{\tau}^{2/3}+ l^{2/3}_{1 \tau}) + \delta^-_k + 2 (k+p)_y l_{1y} + 2l^2_{1y}+ \delta^+_p} \nn\\
&& \frac{1}{-\frac{i c_f}{N} ((l_2 + k)_{\tau}^{2/3}+ l^{2/3}_{2 \tau}) + \delta^-_k + 2 (k+p)_y l_{2y} + 2l^2_{2y}+ \delta^+_p}
\nn\\
&+& \int_0^{\infty} \frac{dk_\tau}{2\pi} \int \frac{d^2 k}{(2 \pi)^2} \int_{k_\tau}^{\infty} \frac{dl_{1 \tau}}{2 \pi} \int_{k_\tau}^{\infty} \frac{dl_{2\tau}}{2 \pi} \int \frac{d l_{1y} dl_{2y}}{(2 \pi)^2} D(l_1) D(l_2) D(l_1-l_2)  \frac{1}{-\frac{i c_f}{N} k^{2/3}_\tau + \delta^-_k}\nn\\
&&\frac{1}{\frac{i c_f}{N} ((l_1 - k)_{\tau}^{2/3}+ l^{2/3}_{1 \tau}) + \delta^-_k + 2 (k+p)_y l_{1y} + 2l^2_{1y} + \delta^+_p}\nn\\ 
&&\frac{1}{\frac{i c_f}{N} ((l_2 - k)_{\tau}^{2/3}+ l^{2/3}_{2 \tau}) + \delta^-_k + 2 (k+p)_y l_{2y}  + 2l^2_{2y}+ \delta^+_p}
\bigg] + h.c.
\eea 
Integrating over $k_x$ and shifting $k_y \to k_y - p$,
\bea &&\delta^{3b} \Sigma(p_\tau=0,\vec{p}) = - N \lambda_+ \lambda_- \bigg[2 i \int_0^{\infty} \frac{dk_\tau}{2\pi} \int \frac{d k_y}{2 \pi} \int_{k_\tau}^{\infty} \frac{dl_{1 \tau}}{2 \pi} \int_0^{\infty} \frac{dl_{2\tau}}{2 \pi} \int \frac{d l_{1y} dl_{2y}}{(2 \pi)^2}\nn\\
&& D(l_1) D(l_2) D(l_{1\tau} + l_{2 \tau}, l_{1y} - l_{2y}) \frac{1}{-\frac{i c_f}{N} (k^{2/3}_{\tau} + (l_1 - k)_{\tau}^{2/3}+ l^{2/3}_{1 \tau})  - 2 k_y l_{1y} + \delta^+_p} \nn\\
&& \frac{1}{-\frac{i c_f}{N} ((l_1 -k)^{2/3}_\tau +l^{2/3}_{1\tau} + (l_2 + k)_{\tau}^{2/3}+ l^{2/3}_{2 \tau}) + 2 k_y (l_{2} - l_{1})_y}
\nn\\
&+& i \int_0^{\infty} \frac{dk_\tau}{2\pi} \int \frac{d k_y}{2 \pi} \int_{k_\tau}^{\infty} \frac{dl_{1 \tau}}{2 \pi} \int_{k_\tau}^{\infty} \frac{dl_{2\tau}}{2 \pi} \int \frac{d l_{1y} dl_{2y}}{(2 \pi)^2} D(l_1) D(l_2) D(l_1-l_2)  \nn\\
&&\frac{1}{\frac{i c_f}{N} (k^{2/3}_\tau + (l_1 - k)_{\tau}^{2/3}+ l^{2/3}_{1 \tau})  + 2 k_y l_{1y} - \delta^+_p} 
\frac{1}{\frac{i c_f}{N} (k^{2/3}_\tau+ (l_2 - k)_{\tau}^{2/3}+ l^{2/3}_{2 \tau}) + 2 k_y l_{2y} - \delta^+_p}
\bigg]+ h.c.\nn\\
\eea 

\bea &&\delta^{3c} \Sigma(p_\tau=0,\vec{p}) = - N \lambda_+ \lambda_- \bigg[2 i \int_0^{\infty} \frac{dk_\tau}{2\pi} \int \frac{d k_y}{2 \pi} \int_{k_\tau}^{\infty} \frac{dl_{1 \tau}}{2 \pi} \int_0^{\infty} \frac{dl_{2\tau}}{2 \pi} \int \frac{d l_{1y} dl_{2y}}{(2 \pi)^2}\nn\\
&& D(l_1) D(l_2) D(l_{1\tau} + l_{2 \tau}, l_{1y} - l_{2y}) \frac{1}{-\frac{i c_f}{N} (k^{2/3}_{\tau} + (l_1 - k)_{\tau}^{2/3}+ l^{2/3}_{1 \tau})  - 2 k_y l_{1y} - 2l^2_{1y} - \delta^+_p} \nn\\
&& \frac{1}{-\frac{i c_f}{N} ((l_1 -k)^{2/3}_\tau +l^{2/3}_{1\tau} + (l_2 + k)_{\tau}^{2/3}+ l^{2/3}_{2 \tau}) + 2 k_y (l_{2} - l_{1})_y + 2 (l^2_{2y} - l^2_{1y})}
\nn\\
&+& i \int_0^{\infty} \frac{dk_\tau}{2\pi} \int \frac{d k_y}{2 \pi} \int_{k_\tau}^{\infty} \frac{dl_{1 \tau}}{2 \pi} \int_{k_\tau}^{\infty} \frac{dl_{2\tau}}{2 \pi} \int \frac{d l_{1y} dl_{2y}}{(2 \pi)^2} D(l_1) D(l_2) D(l_1-l_2)  \nn\\
&&\frac{1}{\frac{i c_f}{N} (k^{2/3}_\tau + (l_1 - k)_{\tau}^{2/3}+ l^{2/3}_{1 \tau})  + 2 k_y l_{1y} + 2 l^2_{1y} + \delta^+_p} 
\frac{1}{\frac{i c_f}{N} (k^{2/3}_\tau+ (l_2 - k)_{\tau}^{2/3}+ l^{2/3}_{2 \tau}) + 2 k_y l_{2y} + 2 l^2_{2y} + \delta^+_p}
\bigg]\nn\\
&& + h.c.
\eea 
The integration regions $l_{1y} > 0$ and $l_{1y} < 0$ give the same contribution. So, integrating over $k_y$,
\bea &&\delta^{3b} \Sigma(p_\tau=0,\vec{p}) =  N \lambda_+ \lambda_- \bigg[2 \int_0^{\infty} \frac{dk_\tau}{2\pi} \int_{k_\tau}^{\infty} \frac{dl_{1 \tau}}{2 \pi} \int_0^{\infty} \frac{dl_{2\tau}}{2 \pi} \int_0^{\infty} \frac{d l_{1y}}{2\pi} \int_{l_{1y}}^{\infty} \frac{dl_{2y}}{2 \pi}\nn\\
&& D(l_1) D(l_2) D(l_{1\tau} + l_{2 \tau}, l_{1y} - l_{2y})\nn\\
&&\frac{1}{-\frac{i c_f}{N} ( l_{2y} ((l_1 - k)_{\tau}^{2/3}+ l^{2/3}_{1 \tau} + k^{2/3}_{\tau}) + l_{1y} ((l_2 + k)_{\tau}^{2/3}+ l^{2/3}_{2 \tau} - k^{2/3}_{\tau}))  + (l_{2} - l_{1})_y \delta^+_p}
\nn\\
&+&  \int_0^{\infty} \frac{dk_\tau}{2\pi} \int_{k_\tau}^{\infty} \frac{dl_{1 \tau}}{2 \pi} \int_{k_\tau}^{\infty} \frac{dl_{2\tau}}{2 \pi} \int_0^{\infty} \frac{d l_{1y}}{2\pi} \int_0^{\infty} \frac{dl_{2y}}{2 \pi} D(l_1) D(l_2) D(l_{1\tau}-l_{2\tau}, l_{1y} + l_{2y})  \nn\\
&&\frac{1}{-\frac{i c_f}{N} ( l_{2y} ((l_1 - k)_{\tau}^{2/3}+ l^{2/3}_{1 \tau} + k^{2/3}_{\tau}) + l_{1y} ((l_2 - k)_{\tau}^{2/3}+ l^{2/3}_{2 \tau} + k^{2/3}_{\tau}))  + (l_{1} + l_{2})_y \delta^+_p}\bigg]+ h.c.\nn\\
\label{Sigmab3D}\eea  
\bea &&\delta^{3c} \Sigma(p_\tau=0,\vec{p}) =  N \lambda_+ \lambda_- \bigg[2 \int_0^{\infty} \frac{dk_\tau}{2\pi} \int_{k_\tau}^{\infty} \frac{dl_{1 \tau}}{2 \pi} \int_0^{\infty} \frac{dl_{2\tau}}{2 \pi} \int_0^{\infty} \frac{d l_{1y}}{2\pi} \int_{l_{1y}}^{\infty} \frac{dl_{2y}}{2 \pi}\nn\\
&& D(l_1) D(l_2) D(l_{1\tau} + l_{2 \tau}, l_{1y} - l_{2y})\nn\\
&&\frac{1}{-\frac{i c_f}{N} ( l_{2y} ((l_1 - k)_{\tau}^{2/3}+ l^{2/3}_{1 \tau} + k^{2/3}_{\tau}) + l_{1y} ((l_2 + k)_{\tau}^{2/3}+ l^{2/3}_{2 \tau} - k^{2/3}_{\tau}))  + 2 l_{1y} l_{2y} (l_{2} - l_{1})_y  - (l_2-l_1)_y \delta^+_p}
\nn\\
&+&  \int_0^{\infty} \frac{dk_\tau}{2\pi} \int_{k_\tau}^{\infty} \frac{dl_{1 \tau}}{2 \pi} \int_{k_\tau}^{\infty} \frac{dl_{2\tau}}{2 \pi} \int_0^{\infty} \frac{d l_{1y}}{2\pi} \int_0^{\infty} \frac{dl_{2y}}{2 \pi} D(l_1) D(l_2) D(l_{1\tau}-l_{2\tau}, l_{1y} + l_{2y})  \nn\\
&&\frac{1}{-\frac{i c_f}{N} ( l_{2y} ((l_1 - k)_{\tau}^{2/3}+ l^{2/3}_{1 \tau} + k^{2/3}_{\tau}) + l_{1y} ((l_2 - k)_{\tau}^{2/3}+ l^{2/3}_{2 \tau} + k^{2/3}_{\tau}))  - 2l_{1y} l_{2y} (l_1 + l_2)_y - (l_{1} + l_{2})_y \delta^+_p}\bigg]\nn\\&+& h.c.
\label{Sigmac3D}\eea  

Expanding the self-energy in $\delta^+_p$ and performing a change of variables $l_{1\tau} = k_\tau x_1$, $l_{2 \tau} = k_\tau x_2$, $l_{1y} = (c_b e^2 k_\tau)^{1/3} y_1$, $l_{2y} = (c_b e^2 k_\tau)^{1/3} y_2$,
\bea \delta^{3b} \Sigma_+(p_\tau=0,\vec{p}) &=& \lambda_+ \lambda_-  (J_1 + J_2) \delta^+_p \int_0^{\infty} \frac{d k_\tau}{k_\tau}\label{dktaub}
\\
\delta^{3c} \Sigma_+(p_\tau=0,\vec{p}) &=& \delta^{3c} \Sigma_+(p_\tau=0,\vec{p} = 0) + \lambda_+ \lambda_-  (J_3 + J_4)\delta^+_p \int_0^{\infty} \frac{d k_\tau}{k_\tau}\label{dktauc}
 \eea 
where

\bea J_1 &=& \frac{6}{\pi^2} \int_1^\infty dx_1 \int_0^{\infty} dx_2 \int_0^{\infty} dy_1 \int_{y_1}^{\infty} dy_2 \frac{y_1 y_2 (y_2-y_1)^2}{(x_1 + y^3_1)(x_2 + y^3_2)(x_1 + x_2 + (y_2 - y_1)^3)}\nn\\ &\times&\frac{1}{(y_2((x_1-1)^{2/3}+x^{2/3}_1 + 1) + y_1 ((x_2+1)^{2/3} + x^{2/3}_2 - 1))^2}\label{J1}\\
J_2 &=& \frac{3}{\pi^2} \int_1^\infty dx_1 \int_1^{\infty} dx_2 \int_0^{\infty} dy_1 \int_{0}^{\infty} dy_2 \frac{y_1 y_2 (y_1+y_2)^2}{(x_1 + y^3_1)(x_2 + y^3_2)(|x_1 - x_2| + (y_1 + y_2)^3)}\nn\\
&\times&\frac{1}{(y_2((x_1-1)^{2/3}+x^{2/3}_1 + 1) + y_1 ((x_2-1)^{2/3} + x^{2/3}_2 + 1))^2} \label{J2}\\
J_3 &=& \frac{6}{\pi^2 N^2} \int_1^\infty dx_1 \int_0^{\infty} dx_2 \int_0^{\infty} dy_1 \int_{y_1}^{\infty} dy_2 \frac{y_1 y_2 (y_2-y_1)^2}{(x_1 + y^3_1)(x_2 + y^3_2)(x_1+x_2 + (y_2 - y_1)^3)}\nn\\
&\times&\frac{3y^2_1y^2_2(y_2-y_1)^2 - \frac{1}{N^2} (y_2((x_1-1)^{2/3} + x^{2/3}_1 + 1) + y_1((x_2+1)^{2/3} +x^{2/3}_2 - 1))^2}{(3 y^2_1 y^2_2 (y_2-y_1)^2 + \frac{1}{N^2}(y_2((x_1-1)^{2/3}+x^{2/3}_1 + 1) + y_1 ((x_2+1)^{2/3} + x^{2/3}_2 - 1))^2)^2} \nn\\
\label{J3}\\
J_4 &=& \frac{3}{\pi^2 N^2} \int_1^\infty dx_1 \int_1^{\infty} dx_2 \int_0^{\infty} dy_1 \int_{0}^{\infty} dy_2 \frac{y_1 y_2 (y_1+y_2)^2}{(x_1 + y^3_1)(x_2 + y^3_2)(|x_1-x_2| + (y_1 + y_2)^3)}\nn\\
&\times&\frac{3 y^2_1 y^2_2(y_1+y_2)^2 - \frac{1}{N^2} (y_2((x_1-1)^{2/3} + x^{2/3}_1 + 1) + y_1((x_2-1)^{2/3} +x^{2/3}_2 + 1))^2}{(3 y^2_1 y^2_2 (y_1+y_2)^2 + \frac{1}{N^2}(y_2((x_1-1)^{2/3}+x^{2/3}_1 + 1) + y_1 ((x_2-1)^{2/3} + x^{2/3}_2 + 1))^2)^2}\nn\\
\label{J4}\eea

Cutting off the $UV$ divergence in (\ref{dktaub}), (\ref{dktauc}) at $k_\tau = \Lambda_\tau \sim \Lambda^3_y/e^2$, we obtain to logarithmic accuracy,
\bea \delta^{3b} \Sigma_+(p_\tau=0,\vec{p}) &=& \lambda_+ \lambda_-  (J_1 + J_2) \delta^+_p \log \frac{\Lambda^3_y}{|\delta^+_p|^{3/2}}
\\
\delta^{3c} \Sigma_+(p_\tau=0,\vec{p}) &=& \delta^{3c} \Sigma_+(p_\tau=0,\vec{p} = 0) + \lambda_+ \lambda_-  (J_3 + J_4)\delta^+_p \log \frac{\Lambda^3_y}{|\delta^+_p|^{3/2}}
\eea 
which is equivalent to Eqs. (\ref{SE3loopb}), (\ref{SE3loopc}) with $J_b = 3(J_1 + J_2)$, $J_c = 3(J_3 + J_4)$. Note that $J_1$ and $J_2$ are constants independent of $N$,
\bea J_1 \approx 0.01276 \\
J_2 \approx 0.02264 \eea
On the other hand, the constants $J_3$ and $J_4$ are $N$ dependent. In the large-$N$ limit we can evaluate these constants analytically to leading logarithmic accuracy by setting $N = \infty$ in the integrand.
\beq J_3 \approx \frac{2}{\pi^2 N^2} \int_1^{\infty} dx_1 \int_0^{\infty} dx_2 \int_0^{\infty} dy_1 \int_{y_1}^{\infty} dy_2 \frac{1}{y_1 y_2 (x_1 + y^3_1) (x_2 + y^3_2) (x_1 + x_2 + (y_2 - y_1)^3)} \eeq
The above integral diverges logarithmically when $y_1, y_2, x_2 \to 0$. Hence,
\beq J_3 \approx \frac{2}{\pi^2 N^2} \int_1^{\infty} \frac{dx_1}{x^2_1} \int_0^{1} dx_2 \int_0^1 dy_1 \int_{y_1}^1 dy_2 \frac{1}{y_1 y_2 (x_2 + y^3_2)} \approx  \frac{2}{\pi^2 N^2} \int_0^1 \frac{dy_2}{y_2} \log (y^{-3}_2) \int_0^{y_2} \frac{d y_1}{y_1}  \label{J3ap1}\eeq 
Inspecting the original integral (\ref{J3}), we observe that the logarithmic divergence in (\ref{J3ap1}) is cut-off when 
$y_1 (y_2 - y_1) \sim \frac{1}{N}$. Hence, 
\beq J_3 \approx \frac{2}{\pi^2 N^2} \int_{N^{-\frac12}}^1 \frac{dy_2}{y_2} \log (y^{-3}_2) \int_{(N y_2)^{-1}}^{y_2} \frac{dy_1}{y_1} \approx \frac{1}{4 \pi^2 N^2} \log^3 N \label{J3ap}\eeq
Similarly, 
\bea J_4 &\approx& \frac{1}{\pi^2 N^2} \int_1^\infty dx_1 \int_1^\infty dx_2 \int_0^{\infty} dy_1 \int_0^{\infty} dy_2 \frac{1}{y_1 y_2 (x_1 + y^3_1) (x_2 + y^3_2) (|x_1 - x_2| + (y_1 + y_2)^3)}\nn\\
 &\approx& \frac{4}{\pi^2 N^2} \int_1^{\infty} \frac{dx_1}{x^2_1} \int_0^{1} \frac{dy_2}{y_2} \int_0^{y_2} \frac{dy_1}{y_1} \log((y_1 + y_2)^{-3}) \label{J4ap1}\eea
Inspecting Eq. (\ref{J4}), we see that the logarithmic divergence in (\ref{J4ap1}) is cut-off when $y_1 y_2 \sim \frac{1}{N}$. Writing, $y_1 = y_2 z$,
\beq J_4 \approx -\frac{12}{\pi^2 N^2} \int_{N^{-\frac12}}^1 \frac{dy_2}{y_2} \int^1_{(N y^2_2)^{-1}} \frac{dz}{z} (\log y_2 + \log(1+z)) \approx \frac{1}{2 \pi^2 N^2} \log^3 N \label{J4ap} \eeq
We note that expressions (\ref{J3ap}), (\ref{J4ap}) do not include subleading polynomial corrections in $\log N$. We can also calculate the constants $J_3$, $J_4$ numerically for $N = 2$,
\bea J_3 \approx -0.004491\\
J_4 \approx -0.008158\eea

Finally, we compute the insertion of the $\phi^2$ operator into the fermion two-point function, which determines the renormalization of the chemical potential $\delta$ away from criticality. The $UV$ contribution at three loop order comes from the diagrams in Figs. \ref{FigSE3loop} b) c) and can be obtained by expanding the bosonic propagators in Eqs. (\ref{Sigmab3D}), (\ref{Sigmac3D}) to linear order in $r$. This yields,
\bea \delta^{3b} \frac{\d \Sigma}{\d r} &\stackrel{UV}{=}&  - N \bigg[2 \int_0^{\infty} \frac{dk_\tau}{2\pi} \int_{k_\tau}^{\infty} \frac{dl_{1 \tau}}{2 \pi} \int_0^{\infty} \frac{dl_{2\tau}}{2 \pi} \int_0^{\infty} \frac{d l_{1y}}{2\pi} \int_{l_{1y}}^{\infty} \frac{dl_{2y}}{2 \pi}\nn\\
&&(D(l_1) + D(l_2) + D(l_{1\tau} + l_{2 \tau}, l_{1y} - l_{2y})) D(l_1) D(l_2) D(l_{1\tau} + l_{2 \tau}, l_{1y} - l_{2y})\nn\\
&&\frac{1}{-\frac{i c_f}{N} ( l_{2y} ((l_1 - k)_{\tau}^{2/3}+ l^{2/3}_{1 \tau} + k^{2/3}_{\tau}) + l_{1y} ((l_2 + k)_{\tau}^{2/3}+ l^{2/3}_{2 \tau} - k^{2/3}_{\tau}))}
\nn\\
&+&  \int_0^{\infty} \frac{dk_\tau}{2\pi} \int_{k_\tau}^{\infty} \frac{dl_{1 \tau}}{2 \pi} \int_{k_\tau}^{\infty} \frac{dl_{2\tau}}{2 \pi} \int_0^{\infty} \frac{d l_{1y}}{2\pi} \int_0^{\infty} \frac{dl_{2y}}{2 \pi}
\nn\\ && (D(l_1)+D(l_2) + D(l_{1\tau}-l_{2\tau}, l_{1y} + l_{2y})) D(l_1) D(l_2) D(l_{1\tau}-l_{2\tau}, l_{1y} + l_{2y})  \nn\\
&&\frac{1}{-\frac{i c_f}{N} ( l_{2y} ((l_1 - k)_{\tau}^{2/3}+ l^{2/3}_{1 \tau} + k^{2/3}_{\tau}) + l_{1y} ((l_2 - k)_{\tau}^{2/3}+ l^{2/3}_{2 \tau} + k^{2/3}_{\tau})) }\bigg]+ h.c.\nn\\
\eea  
\bea &&\delta^{3c} \frac{\d \Sigma}{\d r} \stackrel{UV}{=}  - N  \bigg[2 \int_0^{\infty} \frac{dk_\tau}{2\pi} \int_{k_\tau}^{\infty} \frac{dl_{1 \tau}}{2 \pi} \int_0^{\infty} \frac{dl_{2\tau}}{2 \pi} \int_0^{\infty} \frac{d l_{1y}}{2\pi} \int_{l_{1y}}^{\infty} \frac{dl_{2y}}{2 \pi}\nn\\
&& (D(l_1) + D(l_2) + D(l_{1\tau} + l_{2 \tau}, l_{1y} - l_{2y})) D(l_1) D(l_2) D(l_{1\tau} + l_{2 \tau}, l_{1y} - l_{2y})\nn\\
&&\frac{1}{-\frac{i c_f}{N} ( l_{2y} ((l_1 - k)_{\tau}^{2/3}+ l^{2/3}_{1 \tau} + k^{2/3}_{\tau}) + l_{1y} ((l_2 + k)_{\tau}^{2/3}+ l^{2/3}_{2 \tau} - k^{2/3}_{\tau}))  + 2 l_{1y} l_{2y} (l_{2} - l_{1})_y }
\nn\\
&+&  \int_0^{\infty} \frac{dk_\tau}{2\pi} \int_{k_\tau}^{\infty} \frac{dl_{1 \tau}}{2 \pi} \int_{k_\tau}^{\infty} \frac{dl_{2\tau}}{2 \pi} \int_0^{\infty} \frac{d l_{1y}}{2\pi} \int_0^{\infty} \frac{dl_{2y}}{2 \pi}\\
&&(D(l_1) + D(l_2) + D(l_{1\tau}-l_{2\tau}, l_{1y} + l_{2y})) D(l_1) D(l_2) D(l_{1\tau}-l_{2\tau}, l_{1y} + l_{2y})  \nn\\
&&\frac{1}{-\frac{i c_f}{N} ( l_{2y} ((l_1 - k)_{\tau}^{2/3}+ l^{2/3}_{1 \tau} + k^{2/3}_{\tau}) + l_{1y} ((l_2 - k)_{\tau}^{2/3}+ l^{2/3}_{2 \tau} + k^{2/3}_{\tau}))  - 2l_{1y} l_{2y} (l_1 + l_2)_y }\bigg]\nn\\&+& h.c.
\eea 
We observe that the contribution from the diagram in Fig. \ref{FigSE3loop} b) vanishes, while the diagram in Fig. \ref{FigSE3loop} c) gives upon switching to dimensionless variables,
\beq \delta^{3} \frac{\d \Sigma}{\d r} \stackrel{UV}{=} J_r e^2 \log \Lambda_y \eeq
with
\bea J_r &=& -\frac{36}{\pi^2 N^2} \int_1^\infty dx_1 \int_0^{\infty} dx_2 \int_0^{\infty} dy_1 \int_{y_1}^{\infty} dy_2 \frac{y^2_1 y^2_2 (y_2-y_1)^2}{(x_1 + y^3_1)(x_2 + y^3_2) (x_1 + x_2 + (y_2 - y_1)^3)} \nn
\\
&& \left(\frac{y_1}{x_1 + y^3_1} + \frac{y_2}{x_2 + y^3_2} + \frac{y_2 - y_1}{x_1 + x_2 + (y_2 - y_1)^3}\right) \nn\\
&& \frac{1}{3 y^2_1 y^2_2 (y_2 - y_1)^2 + \frac{1}{N^2} (y_2((x_1-1)^{2/3} + x^{2/3}_1 + 1) + y_1 ((x_2+1)^{2/3} + x^{2/3}_2 - 1))^2}\nn\\
&+& \frac{18}{\pi^2 N^2} \int_1^\infty dx_1 \int_1^{\infty} dx_2 \int_0^{\infty} dy_1 \int_{0}^{\infty} dy_2 \frac{y^2_1 y^2_2 (y_1+y_2)^2}{(x_1 + y^3_1)(x_2 + y^3_2) (|x_1 - x_2| + (y_1 + y_2)^3)} \nn
\\
&& \left(\frac{y_1}{x_1 + y^3_1} + \frac{y_2}{x_2 + y^3_2} + \frac{y_1 + y_2}{|x_1 - x_2| + (y_1 + y_2)^3}\right) \nn\\
&& \frac{1}{3 y^2_1 y^2_2 (y_1 + y_2)^2 + \frac{1}{N^2} (y_2((x_1-1)^{2/3} + x^{2/3}_1 + 1) + y_1 ((x_2-1)^{2/3} + x^{2/3}_2 + 1))^2}\nn\\\eea
Evaluating the above integral, we obtain Eq. (\ref{Jrnum}).

\end{document}